\documentclass[sigconf]{acmart}

\AtBeginDocument{%
  \providecommand\BibTeX{{%
    \normalfont B\kern-0.5em{\scshape i\kern-0.25em b}\kern-0.8em\TeX}}}

\copyrightyear{2022}
\acmYear{2022}
\setcopyright{acmlicensed}\acmConference[MM '22]{Proceedings of the 30th ACM International Conference on Multimedia}{October 10--14, 2022}{Lisboa, Portugal}
\acmBooktitle{Proceedings of the 30th ACM International Conference on Multimedia (MM '22), October 10--14, 2022, Lisboa, Portugal}
\acmPrice{15.00}
\acmDOI{10.1145/3503161.3548253}
\acmISBN{978-1-4503-9203-7/22/10}

\usepackage{subfig}
\usepackage{multirow}
\usepackage{balance}
\newcommand{\tabfig}[2]{\parbox[c]{1.5em}{\includegraphics[width=#1 in]{#2}}}
\usepackage{tabularx}
\usepackage{array}
\newcolumntype{L}[1]{>{\raggedright\let\newline\\\arraybackslash\hspace{0pt}}m{#1}}
\newcolumntype{C}[1]{>{\centering\let\newline\\\arraybackslash\hspace{0pt}}m{#1}}
\newcolumntype{R}[1]{>{\raggedleft\let\newline\\\arraybackslash\hspace{0pt}}m{#1}}

\begin{document}

\title{MESH2IR: Neural Acoustic Impulse Response Generator for Complex 3D Scenes}


\author{Anton Ratnarajah}
\affiliation{%
  \institution{University of Maryland}
  \city{College Park}
  \country{USA}}
\email{jeran@umd.edu}

\author{Zhenyu Tang}
\affiliation{%
  \institution{University of Maryland}
  \city{College Park}
  \country{USA}}
\email{zhy@cs.umd.edu}

\author{Rohith Aralikatti}
\affiliation{%
  \institution{University of Maryland}
  \city{College Park}
  \country{USA}}
\email{rohithca@umd.edu}

\author{Dinesh Manocha}
\affiliation{%
  \institution{University of Maryland}
  \city{College Park}
  \country{USA}}
\email{dmanocha@umd.edu}



\begin{abstract}
We propose a mesh-based neural network (MESH2IR) to generate acoustic impulse responses (IRs) for indoor 3D scenes represented using a mesh.  The IRs are used to create a high-quality sound experience in interactive applications and audio processing. Our method can handle input triangular meshes with arbitrary topologies (2K - 3M triangles). We present a novel training technique to train MESH2IR using energy decay relief and highlight its benefits. We also show that training MESH2IR on IRs preprocessed using our proposed technique significantly improves the accuracy of IR generation. We reduce the non-linearity in the mesh space by transforming 3D scene meshes to latent space using a graph convolution network. Our MESH2IR is more than 200 times faster than a geometric acoustic algorithm on a CPU and can generate more than 10,000 IRs per second on an NVIDIA GeForce RTX 2080 Ti GPU for a given furnished indoor 3D scene. The acoustic metrics are used to characterize the acoustic environment. We show that the acoustic metrics of the IRs predicted from our MESH2IR match the ground truth with less than 10\% error. We also highlight the benefits of MESH2IR on audio and speech processing applications such as speech dereverberation and speech separation. To the best of our knowledge, ours is the first neural-network-based approach to predict IRs from a given 3D scene mesh in real-time.
\end{abstract}



\maketitle

\section{INTRODUCTION}
Rapid developments in interactive applications (e.g., games, virtual environments, speech recognition, etc.) demand realistic sound effects in complex dynamic indoor environments with multiple sound sources. Generating realistic sound effects is still a challenging problem because of the geometric and aural complexity of the real-world environment. The geometric complexity is a measure of the number of vertices and faces needed to represent the complex environment as a 3D mesh. The aural complexity depends on the number of sound sources, the number of static and dynamic objects in the environment and the acoustic effects (e.g., early reflection, late reverberation, diffraction, scattering, etc.). The way the sound propagates from a sound source to the listener can be modeled as an impulse response (IR)~\cite{intro:soundprop}. We can generate sound effects by convolving the IR with a dry sound. The IRs are used to generate plausible sound effects in many interactive applications used for games and AR/VR: Steam Audio~\cite{steam-audio}, Project Acoustics~\cite{projectacoustic}, and Oculus Spatializer~\cite{Oculus-spatializer}.

Recently, many machine learning-based algorithms are proposed to synthesize the sounds in musical instruments~\cite{music1,music2}, estimate the acoustic material properties, and model the acoustic effects from finite objects~\cite{finite4,finite1,finite2,finite3}. Neural-network-based IR generators~\cite{ir-gan,fast-rir} for simple rooms are proposed for speech processing applications (e.g., far-field speech recognition, speech enhancement, speech separation, etc.). FAST-RIR~\cite{fast-rir} is a GAN-based IR generator that takes shoe-box-shaped room dimensions, listener and source positions, and reverberation time as inputs and generates a large IR dataset on the fly. Gaming applications demand fast IR generators for complex scenes. Complex indoor 3D scenes with furniture can be represented in detail using mesh models. Traditional geometric IR simulators take a 3D scene mesh-model and listener and source positions as inputs and generate realistic environmental sound effects~\cite{gsound}. The complexity of geometric IR simulators increases exponentially with the reflection depth~\cite{image-method} and makes them impractical for interactive applications. No current simulation and learning methods can compute real-time IRs for unseen complex dynamic scenes.

Sound propagation can be accurately simulated using wave-based methods. The wave-based methods solve wave equations using different numerical solvers such as the boundary-element method~\cite{boundary-element}, the finite-difference time-domain simulation~\cite{FDTD}, etc. The complexity of the wave-based approach grows as the fourth power of the frequency and is limited to static scenes. Geometric acoustic algorithms are a less complex alternative to the wave-based method. Geometric acoustic algorithms can handle complex environments with dynamic objects~\cite{dynamic1} and multiple sources~\cite{source1}. However, geometric acoustic methods do not model the low-frequency components of the IRs accurately because of the ray assumption. The sound wave can be treated as a ray when the wavelength of the sound is smaller than the obstacles in the environment. At low frequencies under 500 Hz, the ray assumption is not valid for most scenes and results in significant simulation errors. Hybrid sound propagation algorithms~\cite{precompute1,GWA} combine IRs from wave-based and geometric techniques to generate accurate IRs in the human aural range for complex dynamic scenes. Generating IRs corresponding to thousands of sound sources at an interactive rate is not possible with the hybrid method because of its complexity.


\textbf{Main Contributions}: We propose a novel learning-based IR generator (MESH2IR) to generate realistic IRs for furnished 3D scenes with arbitrary topologies (i.e., meshes with 2000 faces to 3 million faces) not seen during the training. For a given complex scene, MESH2IR can generate IRs for any listener and source positions. We perform mesh simplification to even handle complex 3D scenes represented using a mesh with millions of triangles. Our mesh encoder network significantly reduces the input data size by transforming the 3D scene meshes into low-dimensional vectors of a latent space. Mesh simplification and its encoder network allow us to handle general 3D scenes meshes with a varying number of faces. We present an efficient approach to preprocess the IR training dataset and show that training MESH2IR on preprocessed dataset gives a significant improvement in the accuracy of IR generation. We also evaluate the contribution of energy decay relief in improving the IRs generated from MESH2IR. We train our MESH2IR on an IR dataset computed using a hybrid sound propagation algorithm~\cite{GWA}. Therefore, the predicted IRs from MESH2IR exhibit good accuracy for both low-frequency and high-frequency components. MESH2IR can generate more than 10,000 IRs for a given indoor 3D scene. We show that our MESH2IR can generate IRs 200 times faster than a geometric acoustic simulator~\cite{pygsound} on a single CPU. We evaluate the accuracy of the predicted IRs from MESH2IR using power spectrum and acoustic metrics which characterize the acoustic environment. Our MESH2IR can predict the IRs for unseen 3D scenes during training with less than 10\% error in three different acoustic metrics. We also show that far-field speech augmented using the IRs generated from MESH2IR significantly improves the performance in speech processing applications.

\section{RELATED WORKS}
We first give an overview of IR simulators that can compute IRs at interactive rates in Section~\ref{IRsim}. In Section~\ref{IRalgo}, we summarize the deep learning-based algorithms used to predict IRs and describe the prior learning-based IR generator. We highlight various applications in audio and speech processing that can use the predicted IRs in Section~\ref{IRapp}. Finally, we mention different publicly available indoor 3D scenes datasets and give an overview of prior IR datasets. 

\subsection{PHYSICS-BASED IR COMPUTATION} 
\label{IRsim}

Many wave-based, geometric-based, and hybrid interactive IR simulation algorithms have been proposed to simulate IRs for complex scenes~\cite{SURVEY_SOUND}. Wave-based and hybrid algorithms precompute the IRs for a static scene and, at runtime, the IR for an arbitrary listener position is calculated by efficient interpolation techniques~\cite{precompute1,precompute2,precompute3}. These precomputation-based interactive IR simulation algorithms can be used only for static scenes~\cite{SURVEY_SOUND}. Geometric-based algorithms are proposed for interactive IR simulation in dynamic scenes~\cite{dynamic1,dynamic2,dynamic3}. The limitations of geometric-based techniques, such as simulation error at lower frequencies, are inherent in these interactive geometric-based algorithms. There are no general physics-based algorithms known for computing accurate IRs, including low-frequency components, for general dynamic scenes. 

\subsection{LEARNING-BASED IR COMPUTATION}
\label{IRalgo}
Recently, neural-network-based methods have been developed to estimate IRs from the reverberant speech signal~\cite{ir_estimate} or from a single image of the environment~\cite{imageRIR,image2reverb}, to translate synthetic IRs to real-word IRs~\cite{ts-rir}, to interpolate IRs~\cite{deepIR}, and to augment the number of IRs~\cite{augment_IR,ir-gan}. FAST-RIR is a learning-based IR generator that is trained to generate specular and diffuse reflections for a given empty shoe-box-shaped room. FAST-RIR may not compute low-frequency components of IRs accurately.


\subsection{AUDIO PROCESSING USING IR} 
\label{IRapp}
IRs are used in a wide range of practical applications such as audio-visual navigation~\cite{soundspaces,navigation1,neural_acoustic_field}, acoustic matching~\cite{acoustic_matching}, sound rendering~\cite{source1}, sound source localization~\cite{source_localize}, speech enhancement~\cite{speech_enhancement}, speech recognition~\cite{speech_recognition1,speech_recognition2}, and speech separation~\cite{speech_seperate2,speech_seperate1}. In most applications, learning-based networks are trained on large, diverse synthetic datasets and tested in real-world environments~\cite{pygsound}. The performance of the deep neural network trained for a particular application is depended on the similarity of the synthetic training dataset and the real-world test environment~\cite{ir-gan}. Synthetic training datasets are created using IR generators. Therefore, the accuracy of the IR generator plays an important role in practical applications that depends on IRs.

\subsection{INDOOR 3D SCENES \& IR DATABASES}
\label{3Dscene}

The indoor 3D scenes can be either captured using RGB-D videos and reconstructed as meshes (real-world scenes)~\cite{acquired1,acquired2,acquired3} or created by humans using professionally designed software (synthetic models)~\cite{3dfront,designed1}. The mesh quality in the real-world scene datasets is not as good as the quality of the synthetic models because 3D scene reconstruction with accurate geometric details is challenging with existing computer vision methods and capturing hardware. Among the synthetic model datasets, 3D-FRONT~\cite{3dfront} contains large-scale synthetic furnished indoor 3D scenes with fine geometric and texture details. The 3D-FRONT dataset has 6813 CAD houses, where 18,968 rooms are furnished with 13,151 3D furniture objects. The furniture is placed in varying numbers in meaningful locations in each room (e.g., living room, dining room, kitchen, bedroom, etc.)

The publicly available IR datasets are either recorded in a real-world environments (recorded IR)~\cite{realIR1,realIR2,realIR3,realIR4} or generated using IR simulation tools (synthetic IR)~\cite{syntheticIR1,syntheticIR2,soundspaces,GWA}. The recorded IR datasets are limited in size (fewer than 5000 IRs) and number of environments (fewer than 10 rooms), and no sufficient information on the recording conditions is provided to train a deep learning model. Synthetic IR datasets like SoundSpaces~\cite{soundspaces} and GWA~\cite{GWA} contain millions of IRs. The IRs in SoundSpaces is simulated using a geometric acoustic method~\cite{gsound} and GWA dataset contains high-quality IRs computed using a hybrid method.


\begin{figure}[h]
  \centering
  \includegraphics[width=\linewidth]{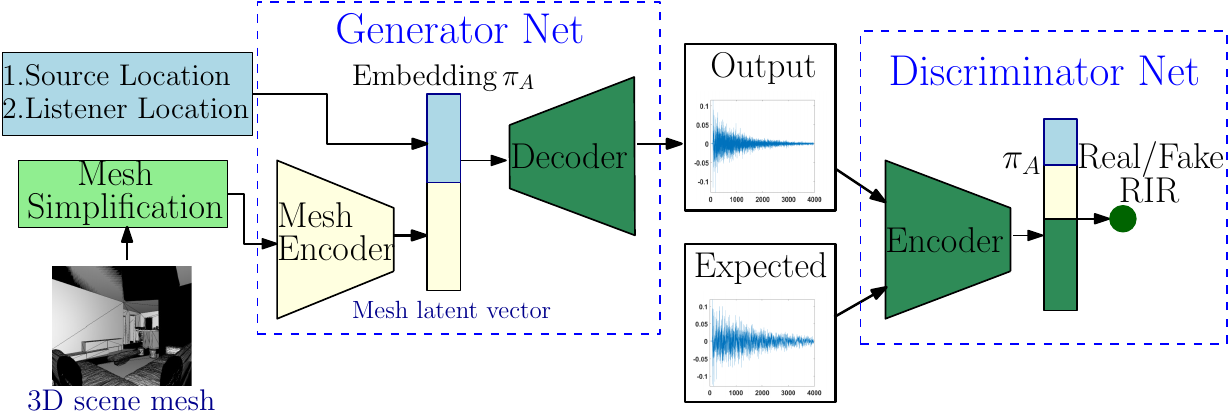}
  \caption{The architecture of our MESH2IR. Our mesh encoder network encodes a indoor 3D scene mesh to the latent space. The mesh latent vector and the source and listener locations are combined to produce a scene vector embedding ($\mathbf{\pi_{A}}$). The generator network generates an IR corresponding to the input scene vector embedding. For the given scene vector embedding, the discriminator network discriminates between the generated IR and the ground truth IR during training.}
  \label{full_architecture}
\end{figure}
\section{MESH2IR: OUR APPROACH}
\label{approach}

\begin{figure*}[h]
  \centering
  \includegraphics[width=\linewidth]{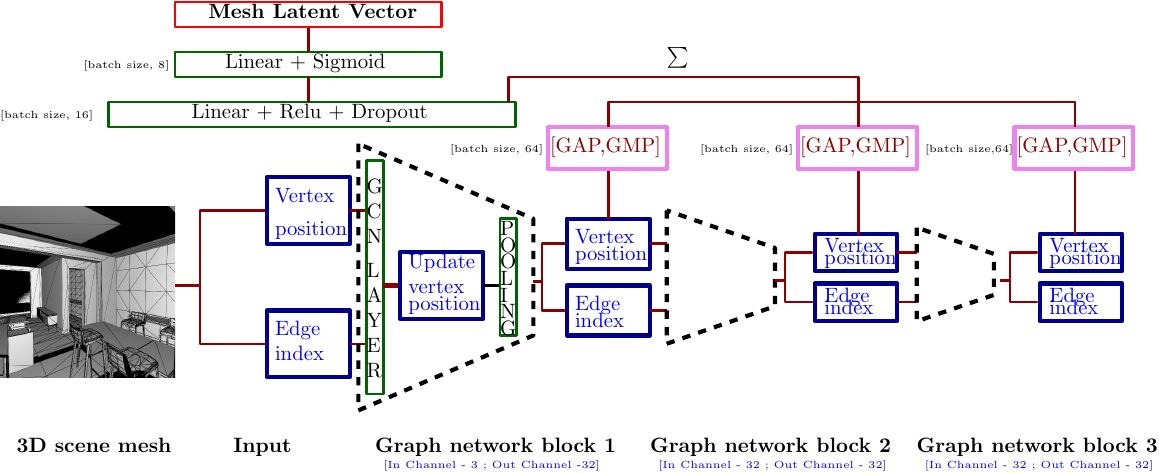} 
  \caption{The expansion of our mesh encoder in Figure~\ref{full_architecture}. Our encoder network transforms the indoor 3D scene mesh into a latent vector. The topology information (edge connectivity) and the node features (vertex coordinates) are extracted from the mesh and passed to our graph neural network.} 
  \label{graph_network}
\end{figure*}
\subsection{OVERVIEW}

Our goal is to predict the IR for the given indoor 3D scene and the source and listener positions (Equation~\ref{overall_network}). The 3D scenes are represented as triangular meshes and we do not make any assumptions about the topology of the 3D scenes. The mesh format describes the shape of an object using vertices, edges, and faces consisting of triangles represented using 3D Cartesian coordinates (x,y,z). The source positions and listener positions are also represented using 3D Cartesian coordinates. For each scene material (e.g., furniture, floor, wall, ceiling, etc.), we do not explicitly control the characteristics of the scene materials (the amount of absorption or scattering of sound by the scene material). The IRs used to train our MESH2IR is computed by considering the characteristic of scene material in every indoor 3D scene. Therefore, our trained MESH2IR randomly assigns the characteristics of scene materials based on their shape.

The IR is the response of an impulse signal emitted in an environment. IR describes the relationship between a dry sound and the reflected sound from the boundaries in the scene. The reflected sound signal depends on the scene geometry, scene materials, and source and listener positions. IRs can be accurately simulated by solving the wave equation~\cite{FDTD}. However, solving a wave equation is not practical for interactive applications because of its complexity. In our work, we propose a learning-based IR generator (MESH2IR) that is capable of approximating thousands of IRs per second for a given complex scene. Our network can be formally described as:
{
\begin{equation}\label{overall_network}
\begin{aligned}[b]
   IR_{n} = \mathbf{N_{\theta_{1}}}(\mathbf{N_{\theta_{2}}}(M_{n}),SP_{n},LP_{n}),
\end{aligned}
\end{equation}
} 
where $IR_{n}$ is the predicted IR for the given scene $n$, $M_{n}$ is the 3D mesh representation of the scene simplified to have around 2000 faces, $LP_{n}$ and $SP_{n}$ are listener location and source location, respectively. $LP_{n}$ and $SP_{n}$ are represented using a 3D vector. We simplify the 3D scene mesh with an arbitrary number of faces to have a constant number of faces (2000 faces). We use a mesh-based encoder network $\mathbf{N_{\theta_{2}}}$ to transform the simplified 3D mesh into a latent vector of a latent space (Figure~\ref{graph_network}). $\mathbf{N_{\theta_{1}}}$ is a generator network, and $\theta_{1}$ and $\theta_{2}$ are the trained network parameters. Our overall network architecture is shown in Figure~\ref{full_architecture}. 

In recent years, cross-modal translation neural networks have gained attention in computer vision. Several algorithms have been proposed for translating video to audio~\cite{video2audio}, image to audio~\cite{image2reverb}, text to image~\cite{stackgan,stackgan++}, image to mesh~\cite{image2mesh}, etc. In our work, we translate complex scenes represented using meshes to acoustic IRs represented as audio signals.

\subsection{TRAINING DATASET}

In our work, we train MESH2IR on the GWA IR dataset~\cite{GWA} simulated using a hybrid algorithm to generate realistic IRs on unseen complex environments. The IRs in GWA are created by automatically calibrating the ray energies simulated using the geometric acoustic method~\cite{pygsound} with the wave effects simulated using finite-difference time-domain wave solver~\cite{FDTD} to create high-quality low-frequency and high-frequency wave effects. The GWA dataset consists of 2 million IRs simulated on the indoor 3D environments represented as meshes in the 3D-FRONT dataset. Out of the 2 million IRs, we train MESH2IR on 200,000 IRs simulated in 5,000 different indoor environments from 3D-FRONT dataset~\cite{3dfront}.

\subsection{MESH SIMPLIFICATION}

The number of faces in the 3D-FRONT dataset varies from about 2000 faces to 3 million faces. We initially simplify the meshes using a quadratic-based edge collapse algorithm in PyMeshLab~\cite{pymeshlab} to have a fewer number of faces (i.e., 2000 faces). The edge collapse algorithm makes sure that the approximation error between the original mesh and the simplified mesh, in terms of Hausdorff distance is small. Therefore the acoustic characteristics do not change due to the simplification step.

\begin{figure}[!h] 
\centering
\subfloat[Before mesh simplification.]{\includegraphics[width=0.45\columnwidth]{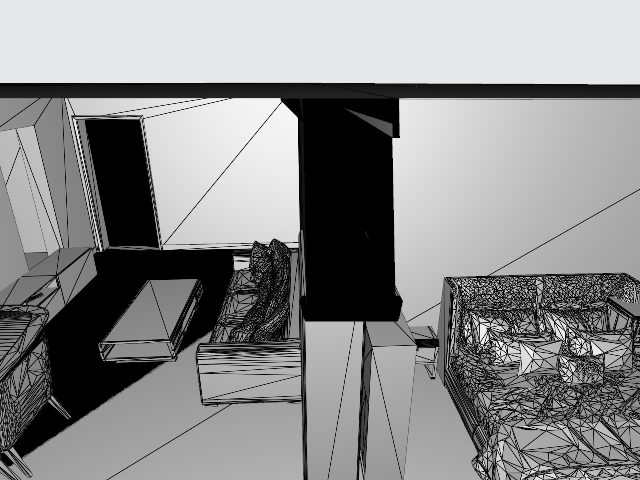}}
\quad
\subfloat[After mesh simplification.]{\includegraphics[width=0.45\columnwidth]{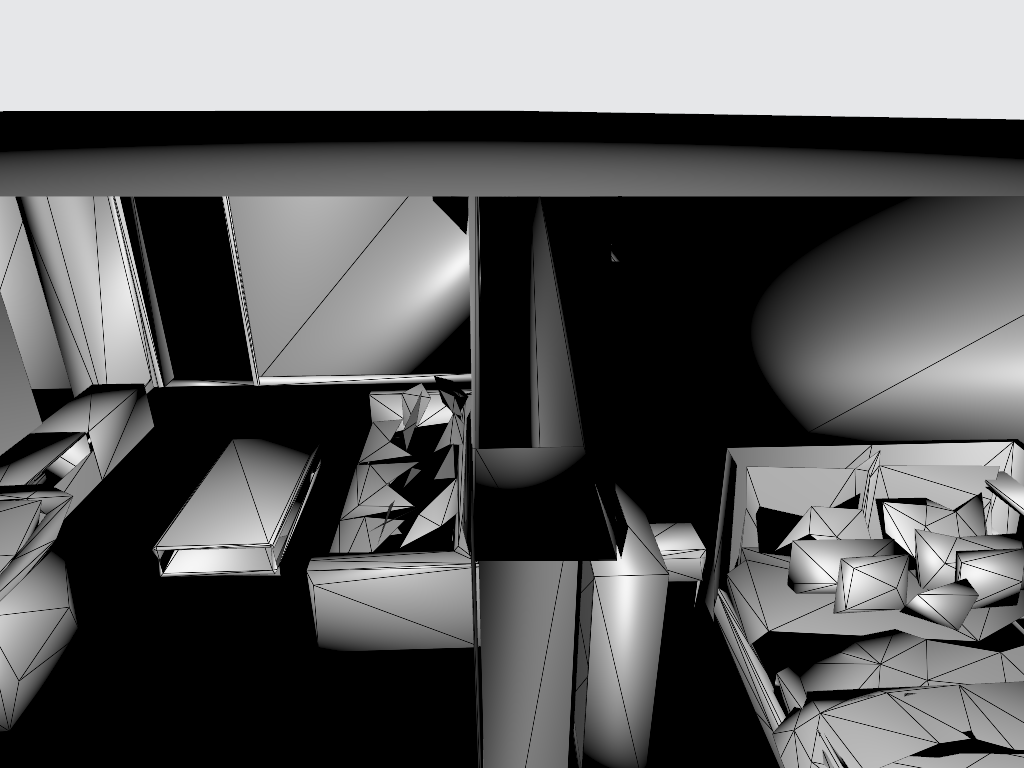}}
\caption{The indoor 3D scene is represented using an original mesh with around 100,000 faces and a simplified mesh with 2,000 faces. We can see that high-level details of the room geometry and furniture are preserved in the simplified mesh. }
\label{mesh_simplification}
\end{figure}

Figure~\ref{mesh_simplification} depicts an example of the original indoor 3D scene mesh and the simplified mesh using the quadratic edge collapse algorithm. We can see that high-level details of the furniture such as bed, pillow, settee, table and cupboard are preserved in the simplified mesh, in addition to the scene geometry. In this example, the simplified mesh has only 2\% of faces of the original mesh. 


\subsection{MESH ENCODER}
Our goal is to transform the simplified indoor scene meshes into a low-dimensional latent space. The triangular meshes can be represented as graph data. Therefore, we represent the 3D scene meshes as a graph and use a graph network (Figure ~\ref{graph_network}) to reduce the dimension. Our graph network uses graph convolution (GCN) layers~\cite{GCN} to encode the topology information (i.e., edge connectivity) and node features (i.e., vertex coordinates) of the graph. We use graph pooling (gpool) layers~\cite{Kpool1,Kpool2} to reduce the topology information of the graph.

The layer-wise propagation rule of multi-layer GCN~\cite{GCN}:
{
\begin{equation}\label{overall_networ3}
\begin{aligned}[b]
   X^{(l+1)} = \sigma(\hat{D}^{-\frac{1}{2}}\hat{A}\hat{D}^{-\frac{1}{2}}X^{(l)}W^{(l)}),
\end{aligned}
\end{equation}
} 
where $\hat{D}_{ii}  = \sum_{j} \hat{A}_{ij}$, and $\hat{A}  = A + I $. $A$ is the adjacency matrix, $I$ is the identity matrix and $W^{(l)}$ is a trainable weight matrix for layer $l$. The feature matrices at layers $l$ and $l+1$ are $X^{(l)}$ and $X^{(l+1)}$, respectively. The topology information is not modified in the GCN layer.

The gpool layer~\cite{Kpool1} creates a new graph with high-level feature encoding by choosing $K$ vertices from the original vertex set and discarding the other vertices. Some edges are removed when discarding vertices in the gpool layer. This results in some isolated vertices in the graph. To address this problem, the gpool algorithm increases the graph connectivity by calculating the square of the adjacency matrix $A^{(l)}$ at layer $l$ and uses the new adjacency matrix $A_n^{(l)}$ for gpool computation (Equation~\ref{gpool1}).
{
\begin{equation}\label{gpool1}
\begin{aligned}[b]
   A_n^{(l)} = A^{(l)} A^{(l)}.
\end{aligned}
\end{equation}
} 

We calculate the channel-wise average (GAP) and the channel-wise maximum (GMP) of the node features after each gpool layer and aggregate the GAP and GMP values. We concatenate the aggregated GAP and aggregated GMP values, pass them to linear layers, and get the mesh latent vector ($\pi_{M}$) (dimension=8) as the output.

\subsection{SCENE VECTOR EMBEDDING}

We concatenate the mesh latent vector ($\pi_{M}$) with the source ($SP$) and listener positions ($LP$) in 3D Cartesian space to generate scene vector embedding $\pi_{A}$ of dimension 14 (Equation~\ref{embedd}). The mesh latent vector will be learned during training. 
{
\begin{equation}\label{embedd}
\begin{aligned}[b]
  \pi_{A}  = [\pi_{M} \; SP \; LP ].
\end{aligned}
\end{equation}
} 

\subsection{IR REPRESENTATION \& PREPROCESSING}
\label{sec:IR_representation}

The IRs in the GWA dataset~\cite{GWA} have a sampling rate of 48 kHz. We downsample the IRs in the dataset to 16 kHz and crop the IRs. Downsampling IRs allows us to maintain a longer duration of IRs in a fixed-length input IR vector (3968 samples). The IRs with full duration and the IRs cropped to have a duration of around 0.25 seconds give similar performance in speech applications~\cite{fast-rir}.

\textbf{IR preprocessing: }The IRs in the GWA dataset have large variations in standard deviation (STD) of the magnitude values ($10^{-12}$ to $10^{-2}$). We noticed that it is hard for MESH2IR to learn from such datasets with high dynamic ranges. To overcome this issue, we divide the IRs by ten times the STD to create preprocessed IRs with the same STD of 0.1 (or any constant STD). We noticed that preprocessing IRs to have constant STD improves the accuracy of MESH2IR (see Section~\ref{ablation:IR_pre}).

To recover the IR with the original magnitude, we duplicate the STD of the IRs 128 times and concatenate them at the end of the preprocessed IRs. The concatenated IRs will have a length of 4096 (3968+128). The convolution layers in MESH2IR calculate an average of 41 neighboring sample values for each sample of the concatenated IRs and pass them to the next layer. Therefore, concatenated STD values near the end of preprocessed IR magnitudes in a concatenated IR will be corrupted when we calculate the average and pass it to the next layers. We can recover the STD values from the tail-end of the concatenated IR where all the 41 neighboring samples are STD.




\subsection{ENERGY DECAY RELIEF}

The energy decay relief (EDR) obtained from the physics-based algorithm contains enough information to be converted into an ``equivalent impulse response''~\cite{energycurve1}. Therefore, we use EDR in the cost function of our MESH2IR. Constraining the Generator network with additional information (i.e., EDR) helps the model to converge easily.

The energy decay curve (EDC) describes the energy remaining in the IR with respect to time~\cite{EDC}. When compared to the IR itself, the EDC decays more smoothly, and we can use it to measure IR acoustic metrics. The generalized EDC for multiple frequency bands is called EDR~\cite{EDR}. EDR is the total amount of energy remaining in the IR at time $t_n$ seconds in a frequency band centered at $f_k$ Hz:

{
\begin{equation}\label{EDRelief}
\begin{aligned}[b]
  EDR(t_n,f_k)  = \sum_{m=n}^M |H(m,k)|^2.
\end{aligned}
\end{equation}
} 

In Equation~\ref{EDRelief}, bin $k$ of the short-time Fourier transform at time-frame $m$ is denoted by $H(m,k)$. $M$ is the total number of time frames. 

\subsection{MODIFIED CONDITIONAL GAN}

Our MESH2IR passes the scene vector embedding $\pi_{A}$ (Equation~\ref{embedd}) to a one-dimensional modified conditional GAN (CGAN) architecture proposed in FAST-RIR~\cite{fast-rir} to generate a single precise IR for the given indoor 3D scene. CGAN~\cite{CGAN1,CGAN2} is conditioned on a random noise $z$ and on a condition $y$ to generate multiple different outputs that satisfy the condition $y$. The modified CGAN is only conditioned on $y$ to generate a single output. 

MESH2IR consists of a generator network ($G_n$) and a discriminator network ($D_n$), which are alternately trained at each iteration. We train $G_n$ and $D_n$ using the IR samples from the data distribution $p_{data}$. The objective function of our generator network consists of modified CGAN error, EDR error, and mean square error (MSE). We train the discriminator network with a modified CGAN objective function. For each $\pi_{A}$, we use the corresponding IRs in the GWA dataset~\cite{GWA} as the ground truth when training our network.

\paragraph{\textbf{Modified CGAN Error (Generator Network):}} The CGAN error is used to generate IRs that are hard for the $D_n$ to differentiate from the ground truth IRs.
{
\begin{equation}\label{CGAN_loss}
\begin{aligned}[b]
    \mathcal{L}_{CGAN} = \mathbb{E}_{\pi_{A} \sim p_{data}}[\log(1 - D_{N}(G_{N}(\pi_{A})))].
\end{aligned}
\end{equation}
} 

\paragraph{\textbf{EDR Error:}} For each $\pi_{A}$, we calculate the EDR of the generated IR using our MESH2IR ($E_N$) and the ground truth IR ($E_G$). We calculate EDR at a set of frequency bands with center frequencies at 125 Hz, 250 Hz, 500 Hz, 1000 Hz, 2000 Hz, and 4000 Hz ($B$). We compare the $E_N$ and $E_G$ for each sample ($s$) as follows:
{
\begin{equation}\label{edr_loss}
\begin{aligned}[b]
    \mathcal{L}_{EDR} = \mathbb{E}_{\pi_{A} \sim p_{data}} [\mathbb{E}_{b \sim B}[\mathbb{E}[(E_{N}(\pi_{A},b,s) - E_{G}(\pi_{A},b,s))^{2}]]].
\end{aligned}
\end{equation}

}

The signal energy in high-frequency components of EDR is high when compared with the signal energy in low-frequency components of EDR in the training IR dataset~\cite{GWA}. Therefore, we give more weight to low-frequency components of EDR.

\paragraph{\textbf{MSE Error: }}
For each $\pi_{A}$, we compare the IR generated from MESH2IR ($I_N$) with the ground truth IR ($I_G$). We calculate the squared difference of each sample ($s$) in $I_N$ and $I_G$ as follows:
{
\begin{equation}\label{mse_loss}
\begin{aligned}[b]
    \mathcal{L}_{MSE} = \mathbb{E}_{\pi_{A} \sim p_{data}}[\mathbb{E}[(I_{N}(\pi_{A},s) - I_{G}(\pi_{A},s))^{2}]].
\end{aligned}
\end{equation}

}

The generator network ($G_N$) and the discriminator network ($D_N$) are trained to compete each other by minimizing the generator objective function $\mathcal{L}_{G_{N}}$ (Equation~\ref{generator_loss})  and maximizing the discriminator objective function $\mathcal{L}_{D_{N}}$ (Equation~\ref{discriminator_loss} ). In Equation~\ref{generator_loss}, the relative importance of the EDR error and MSE error are controlled using $\lambda_{EDR}$ and $\lambda_{MSE}$ respectively.

{
\begin{equation}\label{generator_loss}
\begin{aligned}[b]
    \mathcal{L}_{G_{N}} = \mathcal{L}_{CGAN} + \lambda_{EDR} \; \mathcal{L}_{EDR} + \lambda_{MSE} \; \mathcal{L}_{MSE}.
\end{aligned}
\end{equation}
}

{
\begin{equation}\label{discriminator_loss}
\begin{aligned}[b]
    \mathcal{L}_{D_{N}} = \mathbb{E}_{(I_{G},\pi_{A}) \sim p_{data}}[\log(D_{N}(I_{G}(\pi_{A})))] \\
    + \mathbb{E}_{\pi_{A} \sim p_{data}}[\log(1 - D_{N}(G_{N}(\pi_{A})))].
\end{aligned}
\end{equation}
}

\subsection{IMPLEMENTATION DETAILS}
\paragraph{\textbf{Network Architecture: }} We adapt the graph encoder network in the PyG official repository~\cite{web-pyg} and modify the network to encode indoor 3D scene meshes to the latent space (Figure~\ref{graph_network}). Our gpool layer keeps 60\% of the original number of vertices in each layer. We use the Generator ($G_N$) network and the Discriminator network ($D_N$) architectures proposed in FAST-RIR~\cite{fast-rir} and modify their cost functions. We extend the $G_N$ architecture proposed in FAST-RIR by adding our mesh encoder network (Figure~\ref{full_architecture}).

\paragraph{\textbf{Training:}} We trained $G_N$ and $D_N$ using the RMSprop optimizer with a batch size of 256. The $G_N$ is iterated 3 times for every $D_N$ iteration. Initially, we start with the learning rate of 8x$10^{-5}$ and decay the learning rate by 85\% of the original value for every 7 epochs. We trained our network for 150 epochs. We published our code for reproducibility at github~\cite{web-MESH2IR}.

{
\renewcommand{\arraystretch}{1.5}
\begin{table*}[t]
\centering
\caption{The accuracy of IRs generated using MESH2IR on 5 different furnished indoor 3D scenes not used for training. We plot the time-domain representation of ground truth IRs and the predicted IRs, and evaluate the accuracy of predicted IRs using relative reverberation time ($\mathbf{T_{60}}$) error, relative direct-reverberant-ratio (DRR) error, and relative early-decay-time (EDT) error. Our MESH2IR can predict the IRs with less than 10\% $\mathbf{T_{60}}$ and DRR errors, and less than 3\% EDT error. We also can see that the envelope and magnitude of the predicted IRs matches the ground truth IRs.}
\label{tab:MESH2IR_time_domain}
\begin{tabular}{L{0.1\textwidth}L{0.15\textwidth}L{0.15\textwidth}L{0.15\textwidth}L{0.15\textwidth}L{0.15\textwidth}}
\toprule
Benchmark Scene & \tabfig{1.1600}{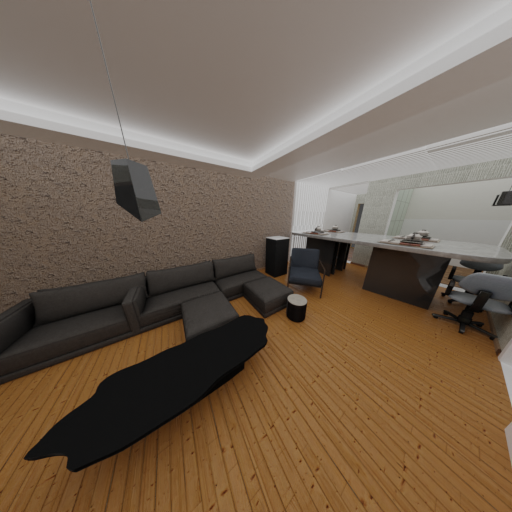}      &  \tabfig{1.1600}{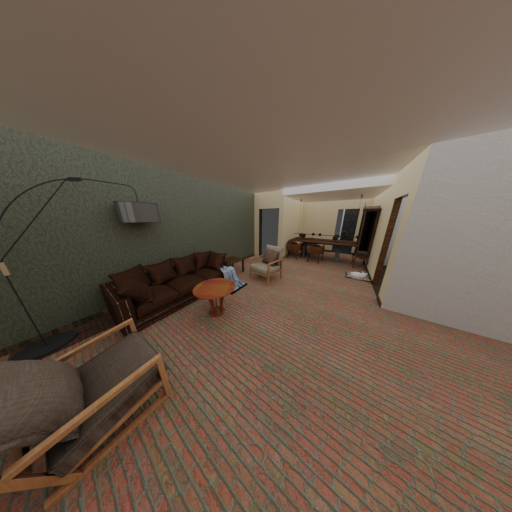}     &    \tabfig{1.1600}{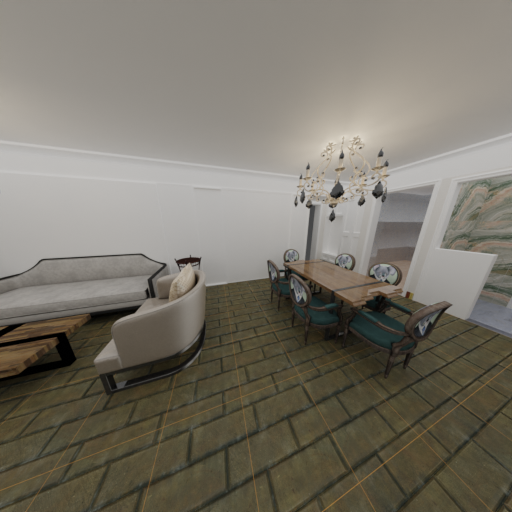}   &    \tabfig{1.1600}{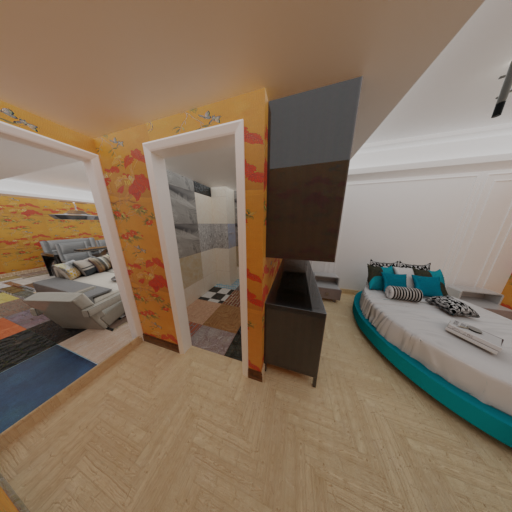}   &   \tabfig{1.1600}{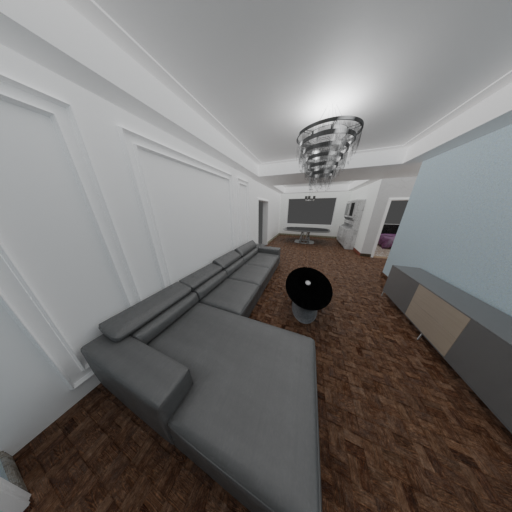}    \\
\# faces              & 185,319     & 92,422     & 181,536     & 253,684     & 282,375     \\
Ground truth IR & 
\scalebox{1.0}[1.00001]{\tabfig{1.1600}{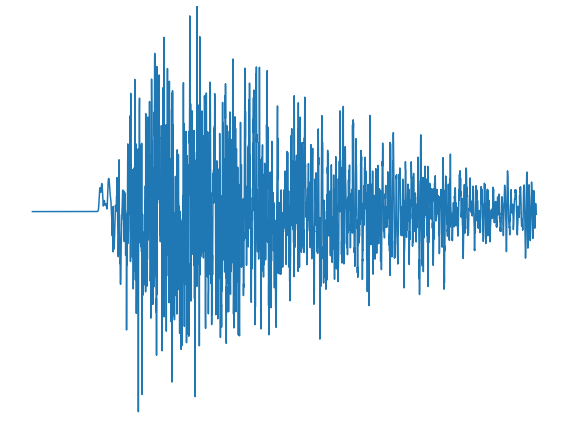}}  & \scalebox{1.0}[1.00001]{\tabfig{1.1600}{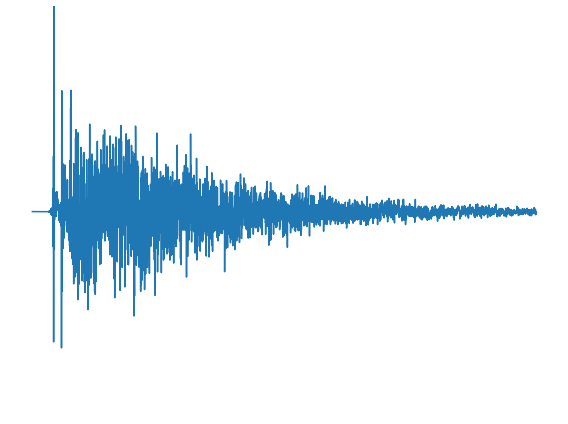}}   & \scalebox{1.0}[1.00001]{\tabfig{1.1600}{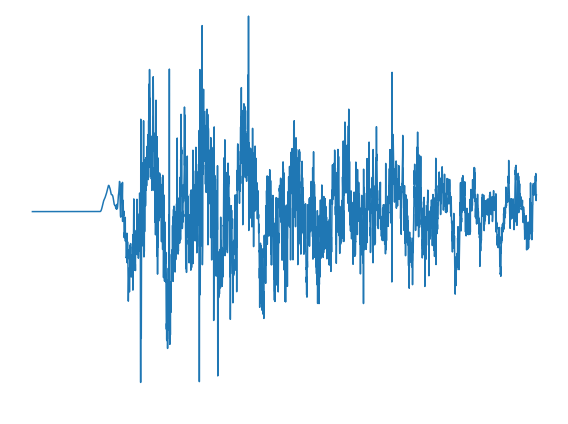}}   & \scalebox{1.0}[1.00001]{\tabfig{1.1600}{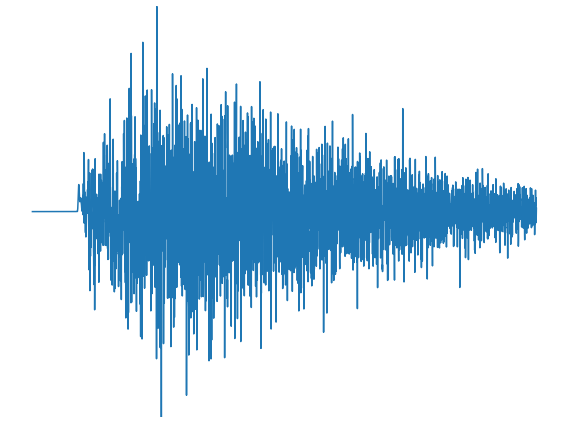}}   & \scalebox{1.0}[1.00001]{\tabfig{1.1600}{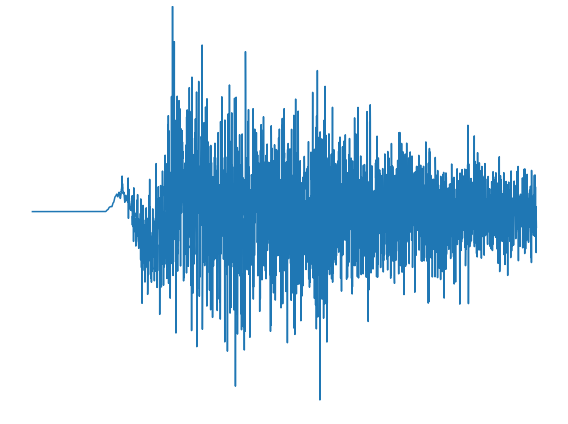}}   \\
Our Mesh2IR & 
\scalebox{1.0}[1.00001]{\tabfig{1.1600}{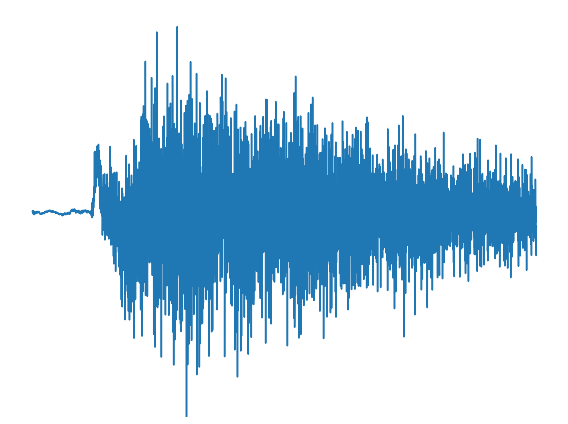}}  & \scalebox{1.0}[1.00001]{\tabfig{1.1600}{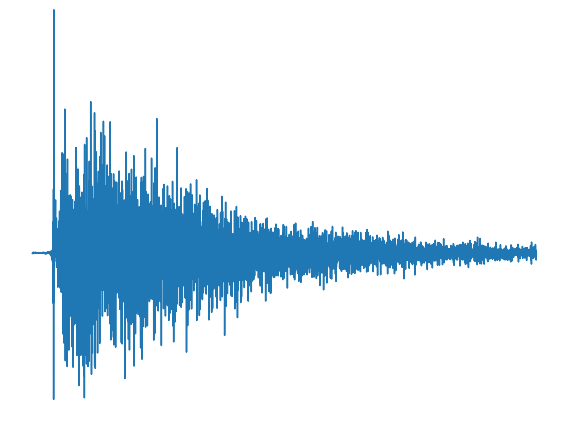}}   & \scalebox{1.0}[1.00001]{\tabfig{1.1600}{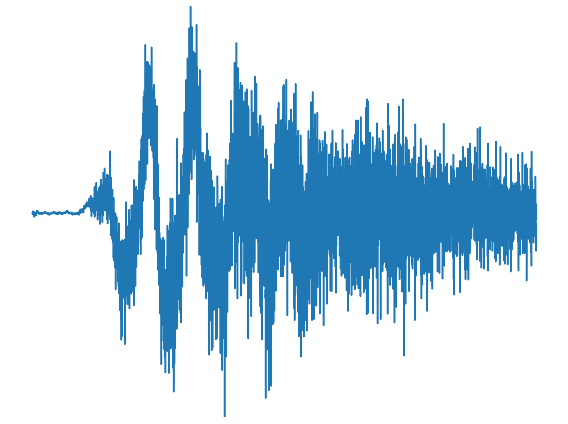}}   & \scalebox{1.0}[1.00001]{\tabfig{1.1600}{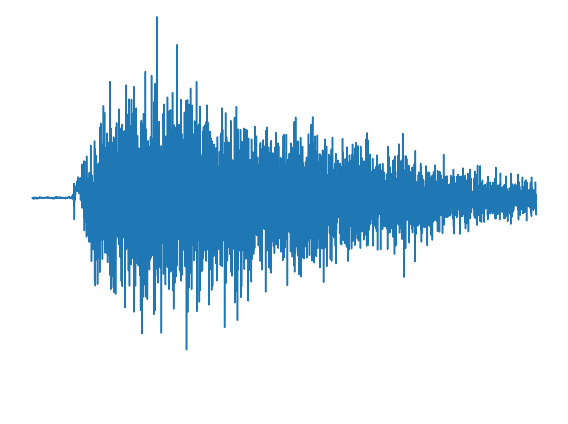}}   & \scalebox{1.0}[1.00001]{\tabfig{1.1600}{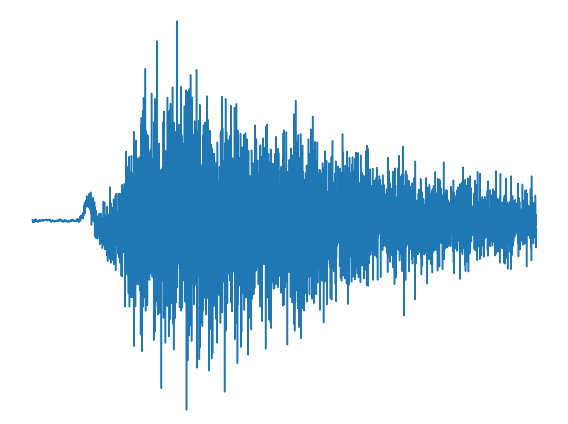}}   \\
$T_{60}$ error &2.7\% &5.9\% &8.2\% &0.6\% &7.5\% \\
DRR error & 2.6\% &9.4\% & 1.4\% &1.9\% &9.3\%\\
EDT error & 2.9\% & 2.1\% & 2.1\% & 1.7\% & 0.9\% \\\bottomrule
\end{tabular}
\end{table*}
}

\section{ABLATION EXPERIMENTS}
\label{ablation}
We perform an ablation study to analyze the importance of our IR preprocessing approach (Section~\ref{sec:IR_representation}) and to find an efficient way of adding EDR to the cost function. We quantitatively evaluate different variations of our network by calculating the mean absolute difference of different acoustic metrics of the generated IRs and the ground truth IRs. The acoustic characteristics of the 3D indoor environment are described using acoustic metrics~\cite{augment_IR}. We also measure the mean square error (MSE) of the generated IRs and the ground truth IRs (Equation~\ref{mse_loss}). We train all the variations of our network with 200K IRs from 5000 different indoor 3D scene meshes and generate 11K testing IRs from 600 scene meshes not used during training.
We use commonly-used acoustic metrics such as reverberation time ($T_{60}$), early-decay-time (EDT), and direct-to-reverberant ratio (DRR) in our experiment. The time taken to decay the sound pressure by 60 decibels (dB) is called $T_{60}$. DRR is the ratio of the sound pressure level of the direct sound to the sound pressure level of reflected sound~\cite{drr_book}. EDT is six times the time taken for the sound source to decay by 10 dB, and it depends on the type and location of the sound source.

\subsection{IR PREPROCESSING}
\label{ablation:IR_pre}
\begin{figure}[!h] 
\centering
\subfloat[Ground truth IR.]{\includegraphics[width=\columnwidth]{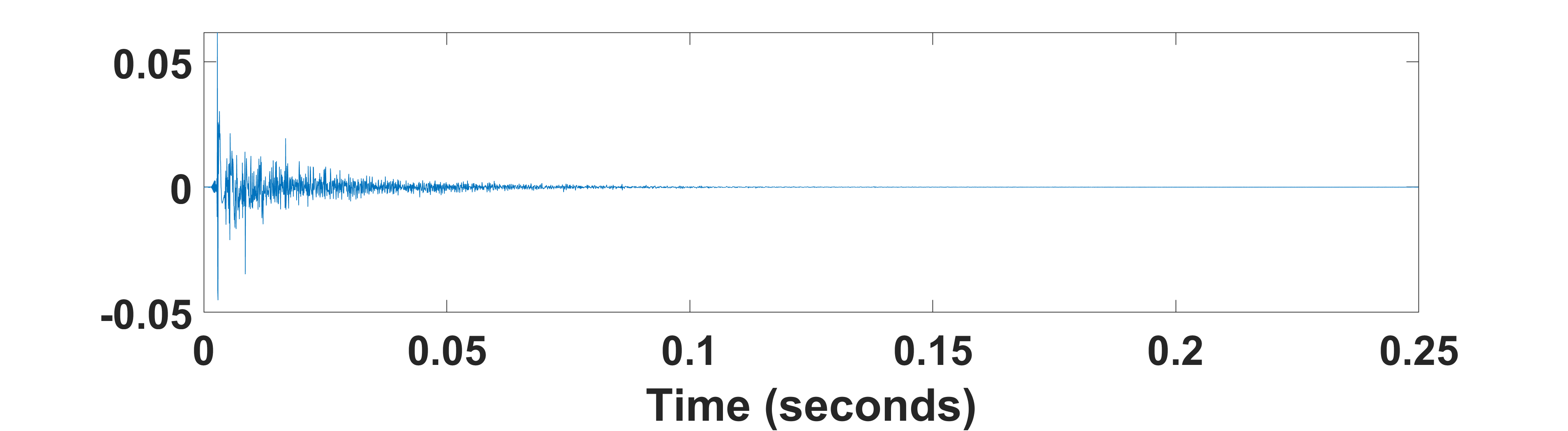}}
\quad
\subfloat[Predicted IR.]{\includegraphics[width=\columnwidth]{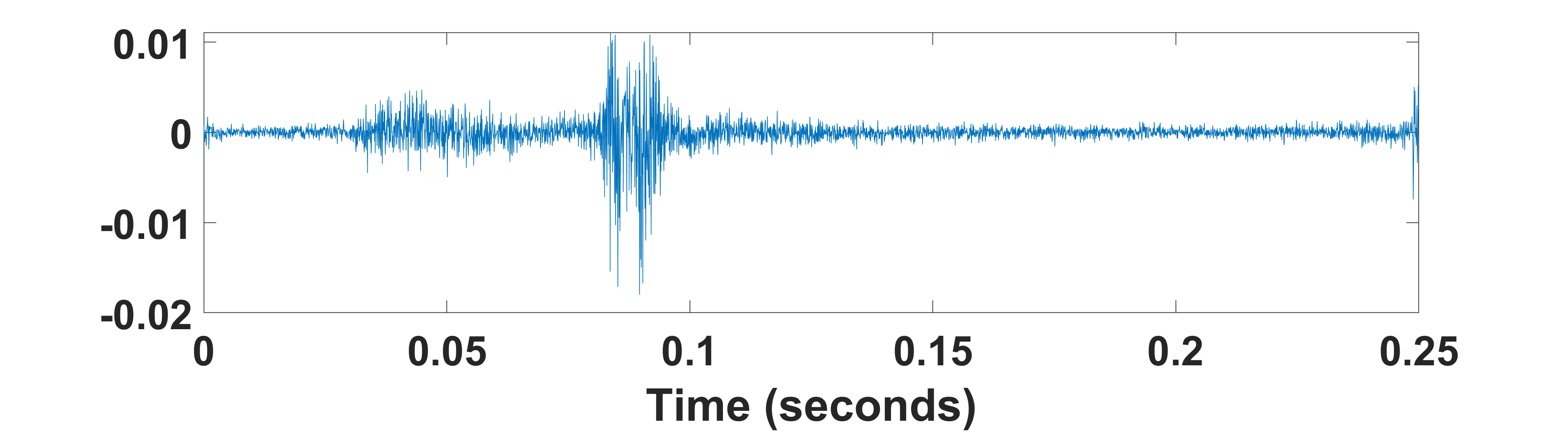}}
\caption{The ground truth IR, and the IR predicted by MESH2IR trained with an unprocessed IR training set (MESH2IR-UNPROCESSED). We can see that MESH2IR-UNPROCESSED gives noisy output.}
\label{ir_pre_process}
\end{figure}

We evaluate the contribution of our IR preprocessing approach discussed in Section~\ref{sec:IR_representation}. We train our MESH2IR on IRs without IR preprocessing (MESH2IR-UNPROCESSED) and generate IRs from our trained network. Figure~\ref{ir_pre_process} shows a ground truth IR and the IR generated from MESH2IR-UNPROCESSED. We can see that MESH2IR-UNPROCESSED generates noisy output. In Table~\ref{tab:MESH2IR_time_domain} we can see that MESH2IR estimates the IRs for the given 3D scene to a greater extent. 

\subsection{ENERGY DECAY RELIEF (EDR)}
\label{ablation:EDR}

To evaluate the importance and the efficient way of adding EDR to the cost functions, we train and compare two different variations of our MESH2IR network.
\newline

\textbf{Variation 1 (MESH2IR-NO-EDR) :} We evaluate the role played by the EDR loss in the generator network by training a variant of MESH2IR without the EDR loss (Equation~\ref{edr_loss}).
\newline

\textbf{Variation 2 (MESH2IR-D-EDR) :} Instead of training the Discriminator network to discriminate between generated IRs and ground truth IRs of dimension 3968 x 1, we train the Discriminator network to discriminate between the EDR of the generated IRs and the ground truth IRs. EDR is calculated at six octave bands with center frequencies at 125 Hz, 250 Hz, 500 Hz, 1000 Hz, 2000 Hz and 4000 Hz. The EDR of an IR has a dimension of 3986x6. We also train the Generator network with the EDR loss.
\newline

From Table~\ref{tab:acoustic_param}, we can see that adding EDR in the generator loss function improves the $T_{60}$, EDT, and MSE. Training the Discriminator network with IRs (MESH2IR) gives a similar performance to passing EDR of the IRs (MESH2IR-D-EDR).

\begin{table}[t]
\caption{\label{tab:acoustic_param} Mean absolute error of the acoustic metrics and mean square error of the generated IRs from MESH2IR-NO-EDR, MESH2IR-D-EDR and MESH2IR. The acoustic metrics used in our experiments are reverberation time ($\mathbf{T_{60}}$) measured in seconds, direct-to-reverberant ratio (DRR) measured in decibels and early-decay-time (EDT) measured in seconds. We can see that overall MESH2IR gives the least error. The best results of each metric are "bolded".}
\begin{tabular}{lcccc}
\hline
\multirow{2}{*}{\textbf{IR Dataset}} & \multicolumn{3}{c}{\textbf{Mean Absolute Error $\downarrow$}} & \multirow{2}{*}{\textbf{MSE (x $\mathbf{10^{-4}})$ $\downarrow$}}                                                                                     \\
\cline{2-4}
& \multicolumn{1}{l}{$\mathbf{T_{60}}$} & \multicolumn{1}{l}{\textbf{DRR}} & \multicolumn{1}{c}{\textbf{EDT}} \\
\hline
MESH2IR-NO-EDR & 0.16 & \textbf{2.68} & 0.23 & 1.44\\
\hline
MESH2IR-D-EDR& \textbf{0.13} & 2.74 & \textbf{0.22} & \textbf{1.23} \\
\hline
\textbf{MESH2IR} & \textbf{0.13} & 2.72  & \textbf{0.22}  & \textbf{1.23} \\
\hline 
\end{tabular}
\end{table}
\subsection{POWER SPECTRUM}
\label{sec:power_spectrum}
\begin{figure}[t] 
\centering
\subfloat[The source is placed around 1m away from the listener.]{\includegraphics[width=\linewidth]{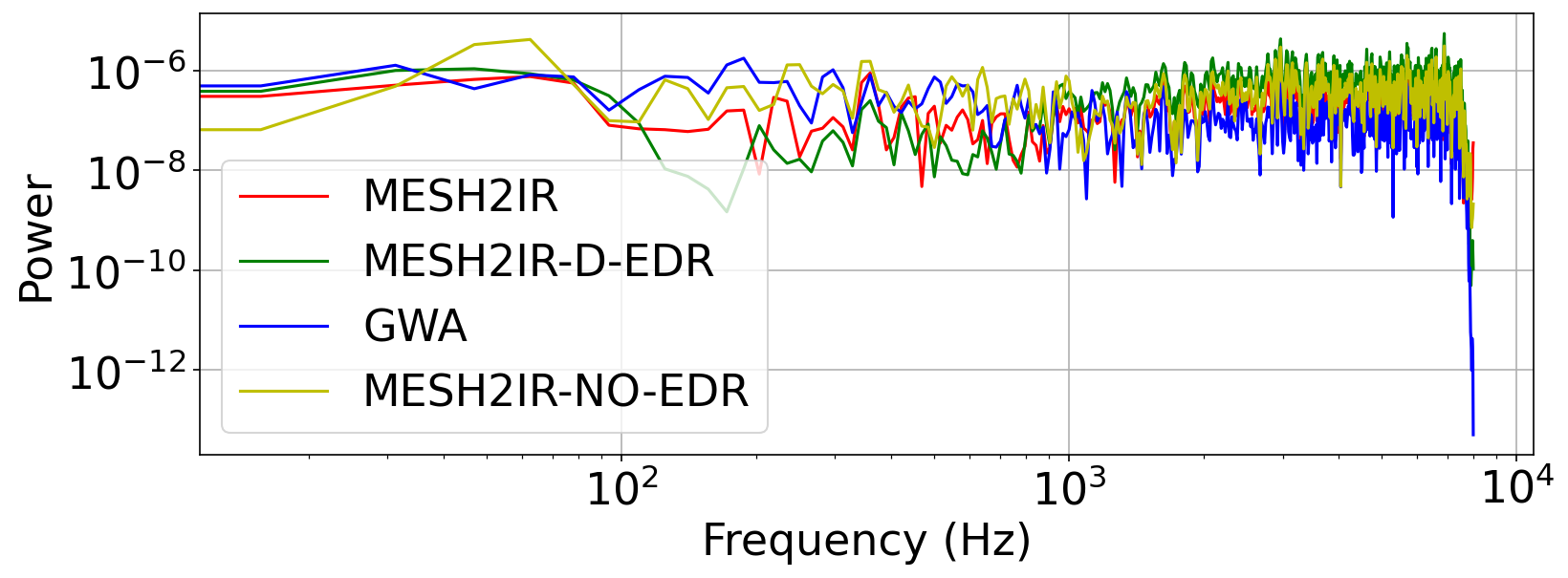}}
\quad
\subfloat[The source is placed around 8m away from the listener.]{\includegraphics[width=\linewidth]{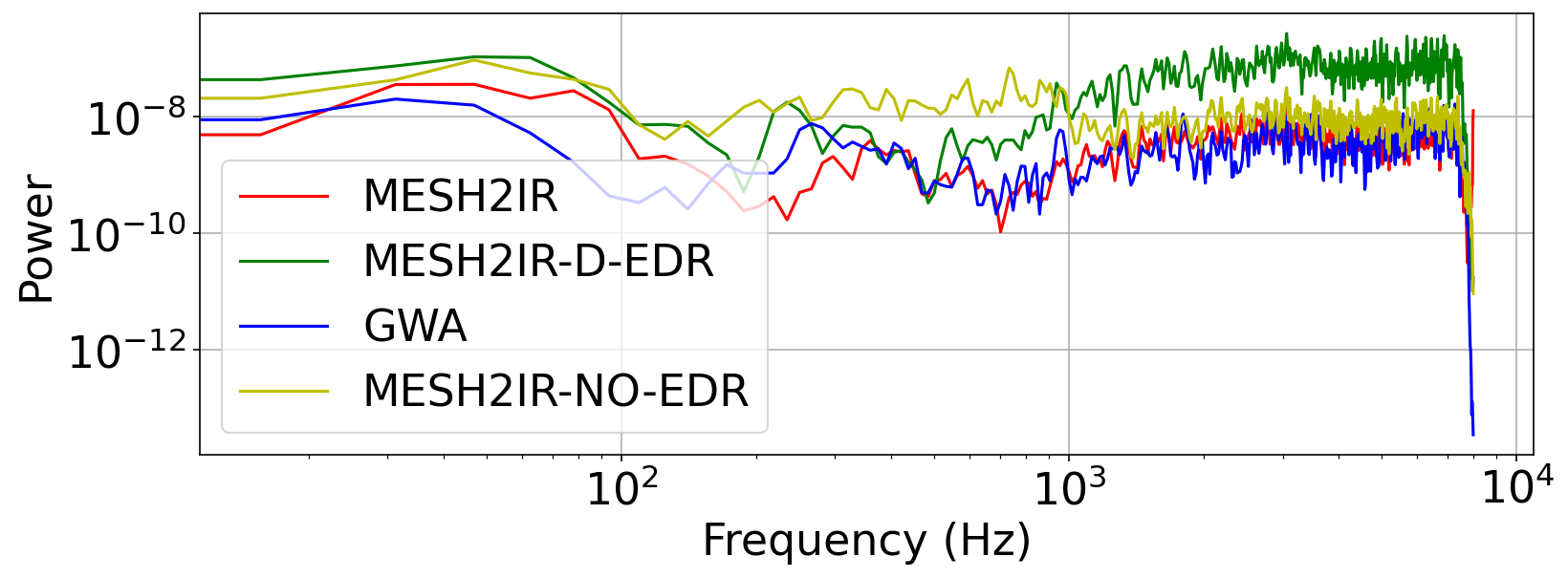}}
\caption{The power spectrum of IRs generated using our proposed MESH2IR, MESH2IR-D-EDR, MESH2IR-NO-EDR, and the ground truth IRs from the GWA dataset. We can see that the power spectrum of MESH2IR is closest to the power spectrum of GWA. }
\label{power_spectrum}
\end{figure}

The power spectrum describes the energy distribution of the IR in the frequency components that compose the waveform. In Figure~\ref{power_spectrum}, we compare the power spectrum of ground truth IRs with the power spectrum of the predicted IR using MESH2IR, MESH2IR-D-EDR, and MESH2IR-NO-EDR. In this example, we placed the listener 1m and 8m away from the source. We can see that in both cases the power spectrum of MESH2IR is closer to GWA when compared with other variants of our approach. 

\section{ACOUSTIC EVALUATION}

In this section, we evaluate the accuracy, power spectrum and runtime of our MESH2IR network. We evaluate the accuracy of our MESH2IR network by comparing the relative acoustic metric error values of the IRs predicted from MESH2IR on indoor 3D scenes not used during training and the ground truth IRs from the GWA dataset~\cite{GWA}. We compare the run-time of MESH2IR network with a state-of-the-art geometric acoustic simulator~\cite{pygsound}.

\subsection{ACCURACY ANALYSIS}

To evaluate the robustness of our MESH2IR, we selected ground truth IRs from the GWA dataset~\cite{GWA} from 5 different indoor 3D scenes with different levels of complexity (the number of faces in a 3D scene mesh). The selected IRs have a large variation in magnitudes and shapes. The distance between the listener and the source position varies from 3.6m to 10.5m. We predicted IRs corresponding to the ground truth IRs in 3D scenes using our MESH2IR. We evaluate the accuracy of our prediction using relative $T_{60}$ error, relative DRR error, and relative EDT error. Table~\ref{tab:MESH2IR_time_domain} shows the ground truth IRs, predicted IRs and the accuracy of our prediction. We can see that our approach can predict the IRs with less than 10\% $T_{60}$ and DRR errors, and less than 3\% EDT error, which are significantly lower than their just-noticeable-differences (JNDs)~\cite{blevins2013quantifying,del2022study,werner2014adjustment}.


\subsection{RUNTIME}
\label{sec:runtime}
We compare the runtime for generating 30,000 IRs from our MESH2IR with the time required to generate these samples from a geometric-based acoustic simulator (GA)~\cite{pygsound} and FAST-RIR~\cite{fast-rir}. The runtime of GA depends on the complexity of the scene, and FAST-RIR is only capable of generating IRs for empty shoe-box-shaped medium-sized rooms (i.e., room dimensions varying from [8m,6m,2.5m]  to [11m,8m,3.5m]). We compare the runtime of GA and FAST-RIR given in Ratnarajah et al. ~\cite{fast-rir} with the runtime of MESH2IR for generating IRs for furnished indoor 3D scene meshes in the 3D-Front dataset~\cite{3dfront}. We use an Intel(R) Xenon(R) CPU E52699 v4 @ 2.20 GHz and a GeForce RTX 2080 Ti GPU for our evaluation. 

For a fair comparison with GA and FAST-RIR methods, which take 3D room dimensions as inputs, we did not consider the time taken for mesh simplification and mesh-to-graph conversion in Table~\ref{tab:runtime}. On average, mesh simplification takes around 7.5 seconds and mesh-to-graph conversion takes 0.04 seconds. From Table~\ref{tab:runtime}, we can see that MESH2IR is more than 200 times faster than GA on a CPU. MESH2IR is optimized to run on GPUs and supports batch parallelization. MESH2IR is slower than FAST-RIR because we encode a 3D scene mesh represented using a graph to a mesh latent vector using our graph neural network (Figure ~\ref{graph_network}). To generate thousands of IRs for a given furnished indoor 3D scene, we perform mesh simplification, mesh-to-graph conversion, and mesh encoding only once.

We have shown the average time taken for mesh encoding (MESH2IR[Mesh Encoder]) and IR generation using the encoded mesh (MESH2IR[IR Generator]) separately in Table~\ref{tab:runtime}. Our MESH2IR can generate more than 10,000 IRs per second on a single GPU for a given furnished indoor 3D scene, and the runtime of MESH2IR is stable irrespective of the scene.

\section{APPLICATIONS}

We demonstrate the benefit of our MESH2IR for several downstream speech applications - neural sound rendering, speech dereverberation and reverberant speech separation. For a fair comparison between different methods, we generate 11K IRs from 600 scene meshes not used during training our MESH2IR using GA~\cite{pygsound}, GWA~\cite{GWA}, MESH2IR-D-EDR, and MESH2IR. We generate reverberant speech using the 11K IRs, and train speech dereverberation and speech separation methods. Reverberant speech can be generated from a dry sound source and an IR via a convolution:
\begin{equation}
x(t) = s(t)*r(t) 
\vspace{-0.1cm}
\end{equation}
where $x(t)$ is the reverberant speech signal, $r(t)$ is the IR, $s(t)$ is the dry speech signal.

\begin{table}[t]
    \setlength{\tabcolsep}{1pt}
	\caption{The runtime of a geometric acoustic simulator (GA)~\cite{pygsound}, FAST-RIR~\cite{fast-rir}, and our MESH2IR. MESH2IR is an extension of FAST-RIR to generate IRs for complex indoor scenes. We can see that the runtime of MESH2IR is higher than FAST-RIR because we use a graph-based network to process the mesh. MESH2IR still outperforms GA on a single CPU.}
	\label{tab:runtime}
	\centering
	\begin{tabular}{@{}lccccr@{}}	
		\toprule
		\textbf{IR Generator} & \textbf{Hardware} &\textbf{Avg time}  & \textbf{Scene Type} \\
		\midrule
		GA \cite{pygsound}& CPU & 30.05s  &Simple\\
		\textbf{MESH2IR} & \textbf{CPU} & \textbf{0.13s}  & \textbf{Complex}\\
		\midrule
		MESH2IR(Batch Size 1) & GPU & 1.32x$10^{-2}$s & Complex\\
		FAST-RIR(Batch Size 128) & GPU & 5.9x$10^{-5}$s & Simple\\ 
		\textbf{MESH2IR(Batch Size 128)} & \textbf{GPU} & \textbf{2.6x\boldmath$10^{-3}$s} & \textbf{Complex}\\
		\midrule
		\textbf{MESH2IR[Mesh Encoder]} & \textbf{GPU} & \textbf{2.57x\boldmath$10^{-3}$s} & \textbf{Complex}\\
		\textbf{MESH2IR[IR Generator]} & \textbf{GPU} & \textbf{7.4x\boldmath$10^{-5}$s} & \textbf{Complex}\\ 
		\bottomrule
	\end{tabular}
\end{table}

\subsection{NEURAL SOUND RENDERING}
Our MESH2IR is the first neural-network-based approach to predict IRs from a given 3D scene mesh at interactive rates. Given this advantage, MESH2IR can be applied to general 3D scenes to enable real-time neural sound rendering, without pre-computation on new scenes. For single IR updates in dynamic scenes, MESH2IR can operate at more than 100 frames per second (fps), which is significantly beyond the requirement for interactive applications (e.g., 10 fps). We demonstrate its sound rendering quality in our supplemental video~\footnote{\url{https://anton-jeran.github.io/M2IR/}}. 

\subsection{SPEECH DEREVERBERATION}
\label{enhancement}


Speech dereverberation is the process of obtaining reverb-free speech from reverberant speech. We train speech dereverberation models using data generated from different synthetic IR generation methods and compare their performance on data generated from recorded IRs. We test the models on IRs from the MIT IR dataset \cite{traer2016statistics}, the BUTReverb dataset \cite{8717722}, and the RWCP Aachen IR dataset \cite{nakamura1999sound, jeub2009binaural}. For all experiments, we train the SkipConvNet~\cite{kothapally2020skipconvnet} model with its default parameters. We use the metric speech-to-reverberation modulation energy ratio (SRMR) \cite{falk2010non} to measure the improvement obtained by the dereverberation process. Higher SRMR implies higher speech quality and lower reverberation effects. We also report the SRMR of the baseline where we do not apply any dereverberation (Reverb).


In Table \ref{tab:realResultsSE}, we test the models on reverberant data generated from recorded IRs. Across all IR datasets, we see that the SRMR of our MESH2IR model is similar to the SRMR of the GWA IRs~\cite{GWA}. We also see that MESH2IR outperforms the GA~\cite{pygsound} method, which only generates IRs using geometric simulations (less accurate than IRs generated from the GWA dataset). The SRMR of MESH2IR is closer to GWA IRs when compared with MESH2IR-D-EDR. Therefore, MESH2IR generates more accurate IRs when compared with MESH2IR-D-EDR.




\begin{table}[t]
\caption{Speech dereverberation results are obtained when training data is generated by different synthetic IR generation methods. Testing is done on reverberant data synthesized from IRs present in three different datasets containing recorded IRs collected in a variety of environments. Higher SRMR is better.}
\label{tab:realResultsSE}
\begin{tabular}{l c c c}
            \hline
\multirow{2}{*}{\textbf{Training Dataset}}             & \multicolumn{3}{c}{\textbf{SRMR}}        \\
             \cline{2-4}
             & \textbf{MIT}                   & \textbf{BUTReverb}                            & \textbf{RWCP Aachen}                    \\
             \hline
Reverb       & 7.35                   & 3.14                         & 5.16                    \\
GA \cite{pygsound} & 6.39 & 3.74   & 4.83  \\
GWA \cite{GWA}    & \textbf{7.67}                   & \textbf{4.6}                 & \textbf{6.14}            \\
\hline
\textbf{MESH2IR-D-EDR}      & 6.18                   & 3.29                            & 4.32                    \\
\textbf{MESH2IR (ours)}       & \textbf{7.82}          & \textbf{4.27}                  & \textbf{5.88}            \\

\hline
\end{tabular}
\end{table}

\subsection{SPEECH SEPARATION}
\label{seperarion}

Speech separation, also referred to as the cocktail party problem, is the process of separating a mixture of speech signals into its constituent speech signals. We check the performance of different synthetic IR generation methods on the task of reverberant speech separation - separating a reverberant mixture into its constituent reverberant speech signals. The dry speech signals and mixtures from the Libri2Mix \cite{cosentino2020librimix} dataset are convolved with IRs to generate the reverberant data. We train the DPRNN-TasNet \cite{luo2020dual} model for all speech separation experiments. We utilize the default implementation provided by the Asteroid \cite{pariente2020asteroid} framework. The model is tested on reverberant mixtures created from real recordings obtained from four different room configurations in the VOiCES \cite{richey2018voices} dataset. We clearly see that our MESH2IR performs similar to GWA~\cite{GWA}, and outperforms GA method~\cite{pygsound} and MESH2IR-D-EDR.

\begin{table}[t]

\caption{Speech separation results in the presence of reverberation are shown below. We report the improvement in the Scale-Invariant Signal Distortion Ratio (SI-SDRi) over the reverberant mixture. Higher Si-SDRi is better. We report performance on reverberant mixtures generated in four different room configurations present in the VOiCES dataset.}
\begin{tabular}{lcccc}
\hline
\multirow{2}{*}{\textbf{Training Dataset}}       & \multicolumn{4}{c}{\textbf{SI-SDRi}}       \\
\cline{2-5}
             & \textbf{Room} 1 & \textbf{Room 2} & \textbf{Room 3} & \textbf{Room 4} \\
\hline

GA \cite{pygsound}   &    2.26    &  2.22       &   1.33     &   2.35     \\
GWA \cite{GWA}     & \textbf{4.75}   & \textbf{4.75}   & \textbf{2.41}   & \textbf{4.91}   \\
\hline
\textbf{MESH2IR-D-EDR}      &  4.68    &   4.35     &   1.87     &   4.72     \\
\textbf{MESH2IR (ours)}       & \textbf{4.91}   & \textbf{4.89}   & \textbf{2.54}   & \textbf{5.13}   \\
\hline
\end{tabular}
\end{table}
\section{CONCLUSION AND FUTURE WORK}
\label{sec:conclusion}
We present a novel neural-network-based MESH2IR architecture to generate thousands of IRs for a given furnished indoor 3D scene on the fly. The IR generation speed of our MESH2IR is constant within a given complex 3D scene, irrespective of the complexity of the scene. Our MESH2IR can generate thousands of IRs per second for a given 3D scene. We show that the IRs predicted by our MESH2IR in unseen indoor 3D scenes are highly similar to the ground truth IRs generated from the GWA dataset, which is used to train our MESH2IR. 

Our work has some limitations. One is we cannot control the characteristics of the scene materials such as the amount of sound absorption and scattering, which can affect the overall amplitude of the IR. Efficiently inputting the characteristics of scene material to our MESH2IR may improve the accuracy of IR generation. 
In addition, while MESH2IR can handle moving sound by simulating many IRs according to a trajectory, when objects in the scene moves, the scene representation changes, which incurs additional encoding time for MESH2IR, making it less efficient in highly dynamic scenes. 
In the future, we would like to integrate our MESH2IR into game engines and other interactive applications, and evaluate its benefits.

\bibliographystyle{ACM-Reference-Format}
\bibliography{sample-base}


\begin{thebibliography}{82}


\ifx \showCODEN    \undefined \def \showCODEN     #1{\unskip}     \fi
\ifx \showDOI      \undefined \def \showDOI       #1{#1}\fi
\ifx \showISBNx    \undefined \def \showISBNx     #1{\unskip}     \fi
\ifx \showISBNxiii \undefined \def \showISBNxiii  #1{\unskip}     \fi
\ifx \showISSN     \undefined \def \showISSN      #1{\unskip}     \fi
\ifx \showLCCN     \undefined \def \showLCCN      #1{\unskip}     \fi
\ifx \shownote     \undefined \def \shownote      #1{#1}          \fi
\ifx \showarticletitle \undefined \def \showarticletitle #1{#1}   \fi
\ifx \showURL      \undefined \def \showURL       {\relax}        \fi
\providecommand\bibfield[2]{#2}
\providecommand\bibinfo[2]{#2}
\providecommand\natexlab[1]{#1}
\providecommand\showeprint[2][]{arXiv:#2}

\bibitem[ste(2018)]%
        {steam-audio}
 \bibinfo{year}{2018}\natexlab{}.
\newblock \bibinfo{booktitle}{\emph{Steam audio}}.
\newblock
\urldef\tempurl%
\url{https://valvesoftware.github.io/steam-audio/}
\showURL{%
\tempurl}


\bibitem[pro(2019)]%
        {projectacoustic}
 \bibinfo{year}{2019}\natexlab{}.
\newblock \bibinfo{booktitle}{\emph{Microsoft project acoustics}}.
\newblock
\urldef\tempurl%
\url{https://docs.microsoft.com/en-us/gaming/acoustics/what-is-acoustics}
\showURL{%
\tempurl}


\bibitem[Ocu(2019)]%
        {Oculus-spatializer}
 \bibinfo{year}{2019}\natexlab{}.
\newblock \bibinfo{booktitle}{\emph{Oculus spatializer}}.
\newblock
\urldef\tempurl%
\url{https://developer.oculus.com/downloads/package/oculus-spatializer-unity/}
\showURL{%
\tempurl}


\bibitem[web(2021)]%
        {web-pyg}
 \bibinfo{year}{2021}\natexlab{}.
\newblock \bibinfo{booktitle}{\emph{PyG official repository : Proteins topk
  pool}}.
\newblock
\urldef\tempurl%
\url{https://github.com/pyg-team/pytorch_geometric/blob/master/examples/proteins_topk_pool.py}
\showURL{%
\tempurl}


\bibitem[web(2022)]%
        {web-MESH2IR}
 \bibinfo{year}{2022}\natexlab{}.
\newblock \bibinfo{booktitle}{\emph{MESH2IR official repository}}.
\newblock
\urldef\tempurl%
\url{https://github.com/anton-jeran/MESH2IR}
\showURL{%
\tempurl}


\bibitem[Allen and Berkley(1979)]%
        {image-method}
\bibfield{author}{\bibinfo{person}{Jont~B. Allen} {and}
  \bibinfo{person}{David~A. Berkley}.} \bibinfo{year}{1979}\natexlab{}.
\newblock \showarticletitle{Image method for efficiently simulating
  small‐room acoustics}.
\newblock \bibinfo{journal}{\emph{The Journal of the Acoustical Society of
  America}} \bibinfo{volume}{65}, \bibinfo{number}{4} (\bibinfo{year}{1979}),
  \bibinfo{pages}{943--950}.
\newblock
\urldef\tempurl%
\url{https://doi.org/10.1121/1.382599}
\showDOI{\tempurl}
\showeprint{https://doi.org/10.1121/1.382599}


\bibitem[Aralikatti et~al\mbox{.}(2021)]%
        {speech_seperate1}
\bibfield{author}{\bibinfo{person}{Rohith Aralikatti}, \bibinfo{person}{Anton
  Ratnarajah}, \bibinfo{person}{Zhenyu Tang}, {and} \bibinfo{person}{Dinesh
  Manocha}.} \bibinfo{year}{2021}\natexlab{}.
\newblock \showarticletitle{Improving Reverberant Speech Separation with
  Synthetic Room Impulse Responses}. In \bibinfo{booktitle}{\emph{{ASRU}}}.
  \bibinfo{publisher}{{IEEE}}, \bibinfo{pages}{900--906}.
\newblock


\bibitem[Avetisyan et~al\mbox{.}(2019)]%
        {acquired3}
\bibfield{author}{\bibinfo{person}{Armen Avetisyan}, \bibinfo{person}{Manuel
  Dahnert}, \bibinfo{person}{Angela Dai}, \bibinfo{person}{Manolis Savva},
  \bibinfo{person}{Angel~X. Chang}, {and} \bibinfo{person}{Matthias
  Nie{\ss}ner}.} \bibinfo{year}{2019}\natexlab{}.
\newblock \showarticletitle{Scan2CAD: Learning {CAD} Model Alignment in {RGB-D}
  Scans}. In \bibinfo{booktitle}{\emph{{CVPR}}}. \bibinfo{publisher}{Computer
  Vision Foundation / {IEEE}}, \bibinfo{pages}{2614--2623}.
\newblock


\bibitem[Blevins et~al\mbox{.}(2013)]%
        {blevins2013quantifying}
\bibfield{author}{\bibinfo{person}{Matthew~G Blevins}, \bibinfo{person}{Adam~T
  Buck}, \bibinfo{person}{Zhao Peng}, {and} \bibinfo{person}{Lily~M Wang}.}
  \bibinfo{year}{2013}\natexlab{}.
\newblock \showarticletitle{Quantifying the just noticeable difference of
  reverberation time with band-limited noise centered around 1000 Hz using a
  transformed up-down adaptive method}.
\newblock  (\bibinfo{year}{2013}).
\newblock


\bibitem[Botteldooren(1995)]%
        {FDTD}
\bibfield{author}{\bibinfo{person}{D. Botteldooren}.}
  \bibinfo{year}{1995}\natexlab{}.
\newblock \showarticletitle{Finite‐difference time‐domain simulation of
  low‐frequency room acoustic problems}.
\newblock \bibinfo{journal}{\emph{The Journal of the Acoustical Society of
  America}} \bibinfo{volume}{98}, \bibinfo{number}{6} (\bibinfo{year}{1995}),
  \bibinfo{pages}{3302--3308}.
\newblock
\urldef\tempurl%
\url{https://doi.org/10.1121/1.413817}
\showDOI{\tempurl}
\showeprint{https://doi.org/10.1121/1.413817}


\bibitem[Bryan(2020)]%
        {augment_IR}
\bibfield{author}{\bibinfo{person}{Nicholas~J. Bryan}.}
  \bibinfo{year}{2020}\natexlab{}.
\newblock \showarticletitle{Impulse Response Data Augmentation and Deep Neural
  Networks for Blind Room Acoustic Parameter Estimation}. In
  \bibinfo{booktitle}{\emph{{ICASSP}}}. \bibinfo{publisher}{{IEEE}},
  \bibinfo{pages}{1--5}.
\newblock


\bibitem[Carlo et~al\mbox{.}(2021)]%
        {realIR4}
\bibfield{author}{\bibinfo{person}{Diego~Di Carlo}, \bibinfo{person}{Pinchas
  Tandeitnik}, \bibinfo{person}{C{\'{e}}dric Foy}, \bibinfo{person}{Antoine
  Deleforge}, \bibinfo{person}{Nancy Bertin}, {and} \bibinfo{person}{Sharon
  Gannot}.} \bibinfo{year}{2021}\natexlab{}.
\newblock \showarticletitle{dEchorate: a Calibrated Room Impulse Response
  Database for Echo-aware Signal Processing}.
\newblock \bibinfo{journal}{\emph{CoRR}}  \bibinfo{volume}{abs/2104.13168}
  (\bibinfo{year}{2021}).
\newblock


\bibitem[Chang et~al\mbox{.}(2017)]%
        {acquired1}
\bibfield{author}{\bibinfo{person}{Angel~X. Chang}, \bibinfo{person}{Angela
  Dai}, \bibinfo{person}{Thomas~A. Funkhouser}, \bibinfo{person}{Maciej
  Halber}, \bibinfo{person}{Matthias Nie{\ss}ner}, \bibinfo{person}{Manolis
  Savva}, \bibinfo{person}{Shuran Song}, \bibinfo{person}{Andy Zeng}, {and}
  \bibinfo{person}{Yinda Zhang}.} \bibinfo{year}{2017}\natexlab{}.
\newblock \showarticletitle{Matterport3D: Learning from {RGB-D} Data in Indoor
  Environments}. In \bibinfo{booktitle}{\emph{3DV}}. \bibinfo{publisher}{{IEEE}
  Computer Society}, \bibinfo{pages}{667--676}.
\newblock


\bibitem[Chen et~al\mbox{.}(2022)]%
        {acoustic_matching}
\bibfield{author}{\bibinfo{person}{Changan Chen}, \bibinfo{person}{Ruohan Gao},
  \bibinfo{person}{Paul Calamia}, {and} \bibinfo{person}{Kristen Grauman}.}
  \bibinfo{year}{2022}\natexlab{}.
\newblock \showarticletitle{Visual Acoustic Matching}.
\newblock \bibinfo{journal}{\emph{CoRR}}  \bibinfo{volume}{abs/2202.06875}
  (\bibinfo{year}{2022}).
\newblock


\bibitem[Chen et~al\mbox{.}(2020)]%
        {soundspaces}
\bibfield{author}{\bibinfo{person}{Changan Chen}, \bibinfo{person}{Unnat Jain},
  \bibinfo{person}{Carl Schissler}, \bibinfo{person}{Sebastia Vicenc~Amengual
  Gari}, \bibinfo{person}{Ziad Al-Halah}, \bibinfo{person}{Vamsi~Krishna
  Ithapu}, \bibinfo{person}{Philip Robinson}, {and} \bibinfo{person}{Kristen
  Grauman}.} \bibinfo{year}{2020}\natexlab{}.
\newblock \showarticletitle{SoundSpaces: Audio-Visual Navigation in 3D
  Environments}. In \bibinfo{booktitle}{\emph{ECCV}}.
\newblock


\bibitem[Chen et~al\mbox{.}(2021)]%
        {navigation1}
\bibfield{author}{\bibinfo{person}{Changan Chen}, \bibinfo{person}{Sagnik
  Majumder}, \bibinfo{person}{Ziad Al{-}Halah}, \bibinfo{person}{Ruohan Gao},
  \bibinfo{person}{Santhosh~Kumar Ramakrishnan}, {and} \bibinfo{person}{Kristen
  Grauman}.} \bibinfo{year}{2021}\natexlab{}.
\newblock \showarticletitle{Learning to Set Waypoints for Audio-Visual
  Navigation}. In \bibinfo{booktitle}{\emph{{ICLR}}}.
  \bibinfo{publisher}{OpenReview.net}.
\newblock


\bibitem[Cosentino et~al\mbox{.}(2020)]%
        {cosentino2020librimix}
\bibfield{author}{\bibinfo{person}{Joris Cosentino}, \bibinfo{person}{Manuel
  Pariente}, \bibinfo{person}{Samuele Cornell}, \bibinfo{person}{Antoine
  Deleforge}, {and} \bibinfo{person}{Emmanuel Vincent}.}
  \bibinfo{year}{2020}\natexlab{}.
\newblock \showarticletitle{Librimix: An open-source dataset for generalizable
  speech separation}.
\newblock \bibinfo{journal}{\emph{arXiv preprint arXiv:2005.11262}}
  (\bibinfo{year}{2020}).
\newblock


\bibitem[Dai et~al\mbox{.}(2017)]%
        {acquired2}
\bibfield{author}{\bibinfo{person}{Angela Dai}, \bibinfo{person}{Angel~X.
  Chang}, \bibinfo{person}{Manolis Savva}, \bibinfo{person}{Maciej Halber},
  \bibinfo{person}{Thomas~A. Funkhouser}, {and} \bibinfo{person}{Matthias
  Nie{\ss}ner}.} \bibinfo{year}{2017}\natexlab{}.
\newblock \showarticletitle{ScanNet: Richly-Annotated 3D Reconstructions of
  Indoor Scenes}. In \bibinfo{booktitle}{\emph{{CVPR}}}.
  \bibinfo{publisher}{{IEEE} Computer Society}, \bibinfo{pages}{2432--2443}.
\newblock


\bibitem[del Solar~Dorrego and Vigeant(2022)]%
        {del2022study}
\bibfield{author}{\bibinfo{person}{Fernando del Solar~Dorrego} {and}
  \bibinfo{person}{Michelle~C Vigeant}.} \bibinfo{year}{2022}\natexlab{}.
\newblock \showarticletitle{A study of the just noticeable difference of early
  decay time for symphonic halls}.
\newblock \bibinfo{journal}{\emph{The Journal of the Acoustical Society of
  America}} \bibinfo{volume}{151}, \bibinfo{number}{1} (\bibinfo{year}{2022}),
  \bibinfo{pages}{80--94}.
\newblock


\bibitem[Falk et~al\mbox{.}(2010)]%
        {falk2010non}
\bibfield{author}{\bibinfo{person}{Tiago~H Falk}, \bibinfo{person}{Chenxi
  Zheng}, {and} \bibinfo{person}{Wai-Yip Chan}.}
  \bibinfo{year}{2010}\natexlab{}.
\newblock \showarticletitle{A non-intrusive quality and intelligibility measure
  of reverberant and dereverberated speech}.
\newblock \bibinfo{journal}{\emph{IEEE Transactions on Audio, Speech, and
  Language Processing}} \bibinfo{volume}{18}, \bibinfo{number}{7}
  (\bibinfo{year}{2010}), \bibinfo{pages}{1766--1774}.
\newblock


\bibitem[Fu et~al\mbox{.}(2021)]%
        {3dfront}
\bibfield{author}{\bibinfo{person}{Huan Fu}, \bibinfo{person}{Bowen Cai},
  \bibinfo{person}{Lin Gao}, \bibinfo{person}{Ling-Xiao Zhang},
  \bibinfo{person}{Jiaming Wang}, \bibinfo{person}{Cao Li},
  \bibinfo{person}{Qixun Zeng}, \bibinfo{person}{Chengyue Sun},
  \bibinfo{person}{Rongfei Jia}, \bibinfo{person}{Binqiang Zhao},
  {et~al\mbox{.}}} \bibinfo{year}{2021}\natexlab{}.
\newblock \showarticletitle{3d-front: 3d furnished rooms with layouts and
  semantics}. In \bibinfo{booktitle}{\emph{Proceedings of the IEEE/CVF
  International Conference on Computer Vision}}. \bibinfo{pages}{10933--10942}.
\newblock


\bibitem[Gao and Ji(2019)]%
        {Kpool1}
\bibfield{author}{\bibinfo{person}{Hongyang Gao} {and}
  \bibinfo{person}{Shuiwang Ji}.} \bibinfo{year}{2019}\natexlab{}.
\newblock \showarticletitle{Graph U-Nets}. In
  \bibinfo{booktitle}{\emph{{ICML}}} \emph{(\bibinfo{series}{Proceedings of
  Machine Learning Research}, Vol.~\bibinfo{volume}{97})}.
  \bibinfo{publisher}{{PMLR}}, \bibinfo{pages}{2083--2092}.
\newblock


\bibitem[Gauthier(2015)]%
        {CGAN2}
\bibfield{author}{\bibinfo{person}{Jon Gauthier}.}
  \bibinfo{year}{2015}\natexlab{}.
\newblock \showarticletitle{Conditional generative adversarial networks for
  convolutional face generation}. In \bibinfo{booktitle}{\emph{Tech Report}}.
\newblock


\bibitem[Grondin et~al\mbox{.}(2020)]%
        {syntheticIR2}
\bibfield{author}{\bibinfo{person}{Fran{\c{c}}ois Grondin},
  \bibinfo{person}{Jean{-}Samuel Lauzon}, \bibinfo{person}{Simon Michaud},
  \bibinfo{person}{Mirco Ravanelli}, {and} \bibinfo{person}{Fran{\c{c}}ois
  Michaud}.} \bibinfo{year}{2020}\natexlab{}.
\newblock \showarticletitle{{BIRD:} Big Impulse Response Dataset}.
\newblock \bibinfo{journal}{\emph{CoRR}}  \bibinfo{volume}{abs/2010.09930}
  (\bibinfo{year}{2020}).
\newblock


\bibitem[Grumiaux et~al\mbox{.}(2021)]%
        {source_localize}
\bibfield{author}{\bibinfo{person}{Pierre{-}Amaury Grumiaux},
  \bibinfo{person}{Srdan Kitic}, \bibinfo{person}{Laurent Girin}, {and}
  \bibinfo{person}{Alexandre Gu{\'{e}}rin}.} \bibinfo{year}{2021}\natexlab{}.
\newblock \showarticletitle{A Survey of Sound Source Localization with Deep
  Learning Methods}.
\newblock \bibinfo{journal}{\emph{CoRR}}  \bibinfo{volume}{abs/2109.03465}
  (\bibinfo{year}{2021}).
\newblock


\bibitem[Hadad et~al\mbox{.}(2014)]%
        {realIR1}
\bibfield{author}{\bibinfo{person}{Elior Hadad}, \bibinfo{person}{Florian
  Heese}, \bibinfo{person}{Peter Vary}, {and} \bibinfo{person}{Sharon Gannot}.}
  \bibinfo{year}{2014}\natexlab{}.
\newblock \showarticletitle{Multichannel audio database in various acoustic
  environments}. In \bibinfo{booktitle}{\emph{{IWAENC}}}.
  \bibinfo{publisher}{{IEEE}}, \bibinfo{pages}{313--317}.
\newblock


\bibitem[Handa et~al\mbox{.}(2015)]%
        {designed1}
\bibfield{author}{\bibinfo{person}{Ankur Handa}, \bibinfo{person}{Viorica
  Patraucean}, \bibinfo{person}{Vijay Badrinarayanan}, \bibinfo{person}{Simon
  Stent}, {and} \bibinfo{person}{Roberto Cipolla}.}
  \bibinfo{year}{2015}\natexlab{}.
\newblock \showarticletitle{SceneNet: Understanding Real World Indoor Scenes
  With Synthetic Data}.
\newblock \bibinfo{journal}{\emph{CoRR}}  \bibinfo{volume}{abs/1511.07041}
  (\bibinfo{year}{2015}).
\newblock


\bibitem[Hawley et~al\mbox{.}(2020)]%
        {music2}
\bibfield{author}{\bibinfo{person}{Scott~H Hawley}, \bibinfo{person}{Vasileios
  Chatziiannou}, {and} \bibinfo{person}{Andrew Morrison}.}
  \bibinfo{year}{2020}\natexlab{}.
\newblock \showarticletitle{Synthesis of musical instrument sounds:
  Physics-based modeling or machine learning}.
\newblock \bibinfo{journal}{\emph{Phys. Today}}  \bibinfo{volume}{16}
  (\bibinfo{year}{2020}), \bibinfo{pages}{20--28}.
\newblock


\bibitem[Jenrungrot et~al\mbox{.}(2020)]%
        {speech_seperate2}
\bibfield{author}{\bibinfo{person}{Teerapat Jenrungrot}, \bibinfo{person}{Vivek
  Jayaram}, \bibinfo{person}{Steven~M. Seitz}, {and} \bibinfo{person}{Ira
  Kemelmacher{-}Shlizerman}.} \bibinfo{year}{2020}\natexlab{}.
\newblock \showarticletitle{The Cone of Silence: Speech Separation by
  Localization}. In \bibinfo{booktitle}{\emph{NeurIPS}}.
\newblock


\bibitem[Jeub et~al\mbox{.}(2009)]%
        {jeub2009binaural}
\bibfield{author}{\bibinfo{person}{Marco Jeub}, \bibinfo{person}{Magnus
  Schafer}, {and} \bibinfo{person}{Peter Vary}.}
  \bibinfo{year}{2009}\natexlab{}.
\newblock \showarticletitle{A binaural room impulse response database for the
  evaluation of dereverberation algorithms}. In \bibinfo{booktitle}{\emph{2009
  16th International Conference on Digital Signal Processing}}. IEEE,
  \bibinfo{pages}{1--5}.
\newblock


\bibitem[Ji et~al\mbox{.}(2020)]%
        {music1}
\bibfield{author}{\bibinfo{person}{Shulei Ji}, \bibinfo{person}{Jing Luo},
  {and} \bibinfo{person}{Xinyu Yang}.} \bibinfo{year}{2020}\natexlab{}.
\newblock \showarticletitle{A Comprehensive Survey on Deep Music Generation:
  Multi-level Representations, Algorithms, Evaluations, and Future Directions}.
\newblock \bibinfo{journal}{\emph{arXiv preprint arXiv:2011.06801}}
  (\bibinfo{year}{2020}).
\newblock


\bibitem[Jin et~al\mbox{.}(2021)]%
        {finite3}
\bibfield{author}{\bibinfo{person}{Xutong Jin}, \bibinfo{person}{Sheng Li},
  \bibinfo{person}{Dinesh Manocha}, {and} \bibinfo{person}{Guoping Wang}.}
  \bibinfo{year}{2021}\natexlab{}.
\newblock \showarticletitle{DeepEigen: Learning-based Modal Sound Synthesis
  with Acoustic Transfer Maps}.
\newblock \bibinfo{journal}{\emph{CoRR}}  \bibinfo{volume}{abs/2108.07425}
  (\bibinfo{year}{2021}).
\newblock


\bibitem[Jin et~al\mbox{.}(2020)]%
        {finite4}
\bibfield{author}{\bibinfo{person}{Xutong Jin}, \bibinfo{person}{Sheng Li},
  \bibinfo{person}{Tianshu Qu}, \bibinfo{person}{Dinesh Manocha}, {and}
  \bibinfo{person}{Guoping Wang}.} \bibinfo{year}{2020}\natexlab{}.
\newblock \showarticletitle{Deep-modal: real-time impact sound synthesis for
  arbitrary shapes}. In \bibinfo{booktitle}{\emph{Proceedings of the 28th ACM
  International Conference on Multimedia}}. \bibinfo{pages}{1171--1179}.
\newblock


\bibitem[Jot(1992)]%
        {EDR}
\bibfield{author}{\bibinfo{person}{Jean{-}Marc Jot}.}
  \bibinfo{year}{1992}\natexlab{}.
\newblock \showarticletitle{An analysis/synthesis approach to real-time
  artificial reverberation}. In \bibinfo{booktitle}{\emph{{ICASSP}}}.
  \bibinfo{publisher}{{IEEE} Computer Society}, \bibinfo{pages}{221--224}.
\newblock


\bibitem[Kipf and Welling(2017)]%
        {GCN}
\bibfield{author}{\bibinfo{person}{Thomas~N. Kipf} {and} \bibinfo{person}{Max
  Welling}.} \bibinfo{year}{2017}\natexlab{}.
\newblock \showarticletitle{Semi-Supervised Classification with Graph
  Convolutional Networks}. In \bibinfo{booktitle}{\emph{{ICLR} (Poster)}}.
  \bibinfo{publisher}{OpenReview.net}.
\newblock


\bibitem[Knyazev et~al\mbox{.}(2019)]%
        {Kpool2}
\bibfield{author}{\bibinfo{person}{Boris Knyazev}, \bibinfo{person}{Graham~W.
  Taylor}, {and} \bibinfo{person}{Mohamed~R. Amer}.}
  \bibinfo{year}{2019}\natexlab{}.
\newblock \showarticletitle{Understanding Attention and Generalization in Graph
  Neural Networks}. In \bibinfo{booktitle}{\emph{NeurIPS}}.
  \bibinfo{pages}{4204--4214}.
\newblock


\bibitem[Ko et~al\mbox{.}(2017)]%
        {syntheticIR1}
\bibfield{author}{\bibinfo{person}{Tom Ko}, \bibinfo{person}{Vijayaditya
  Peddinti}, \bibinfo{person}{Daniel Povey}, \bibinfo{person}{Michael~L.
  Seltzer}, {and} \bibinfo{person}{Sanjeev Khudanpur}.}
  \bibinfo{year}{2017}\natexlab{}.
\newblock \showarticletitle{A study on data augmentation of reverberant speech
  for robust speech recognition}. In \bibinfo{booktitle}{\emph{{ICASSP}}}.
  \bibinfo{publisher}{{IEEE}}, \bibinfo{pages}{5220--5224}.
\newblock


\bibitem[kon and koike(2019)]%
        {imageRIR}
\bibfield{author}{\bibinfo{person}{homare kon} {and} \bibinfo{person}{hideki
  koike}.} \bibinfo{year}{2019}\natexlab{}.
\newblock \showarticletitle{estimation of late reverberation characteristics
  from a single two-dimensional environmental image using convolutional neural
  networks}.
\newblock \bibinfo{journal}{\emph{journal of the audio engineering society}}
  \bibinfo{volume}{67}, \bibinfo{number}{7/8} (\bibinfo{date}{july}
  \bibinfo{year}{2019}), \bibinfo{pages}{540--548}.
\newblock
\urldef\tempurl%
\url{https://doi.org/10.17743/jaes.2018.0069}
\showDOI{\tempurl}


\bibitem[Kothapally et~al\mbox{.}(2020)]%
        {kothapally2020skipconvnet}
\bibfield{author}{\bibinfo{person}{Vinay Kothapally}, \bibinfo{person}{Wei
  Xia}, \bibinfo{person}{Shahram Ghorbani}, \bibinfo{person}{John~HL Hansen},
  \bibinfo{person}{Wei Xue}, {and} \bibinfo{person}{Jing Huang}.}
  \bibinfo{year}{2020}\natexlab{}.
\newblock \showarticletitle{Skipconvnet: Skip convolutional neural network for
  speech dereverberation using optimally smoothed spectral mapping}.
\newblock \bibinfo{journal}{\emph{arXiv preprint arXiv:2007.09131}}
  (\bibinfo{year}{2020}).
\newblock


\bibitem[Koyama et~al\mbox{.}(2021)]%
        {realIR3}
\bibfield{author}{\bibinfo{person}{Shoichi Koyama}, \bibinfo{person}{Tomoya
  Nishida}, \bibinfo{person}{Keisuke Kimura}, \bibinfo{person}{Takumi Abe},
  \bibinfo{person}{Natsuki Ueno}, {and} \bibinfo{person}{Jesper
  Brunnstr{\"{o}}m}.} \bibinfo{year}{2021}\natexlab{}.
\newblock \showarticletitle{{MESHRIR:} {A} Dataset of Room Impulse Responses on
  Meshed Grid Points for Evaluating Sound Field Analysis and Synthesis
  Methods}. In \bibinfo{booktitle}{\emph{{WASPAA}}}.
  \bibinfo{publisher}{{IEEE}}, \bibinfo{pages}{1--5}.
\newblock


\bibitem[Kuttruff(2016)]%
        {intro:soundprop}
\bibfield{author}{\bibinfo{person}{Heinrich Kuttruff}.}
  \bibinfo{year}{2016}\natexlab{}.
\newblock \bibinfo{booktitle}{\emph{Room acoustics}}.
\newblock \bibinfo{publisher}{Crc Press}.
\newblock


\bibitem[kuttruff(1993)]%
        {energycurve1}
\bibfield{author}{\bibinfo{person}{k.~heinrich kuttruff}.}
  \bibinfo{year}{1993}\natexlab{}.
\newblock \showarticletitle{auralization of impulse responses modeled on the
  basis of ray-tracing results}.
\newblock \bibinfo{journal}{\emph{journal of the audio engineering society}}
  \bibinfo{volume}{41}, \bibinfo{number}{11} (\bibinfo{date}{november}
  \bibinfo{year}{1993}), \bibinfo{pages}{876--880}.
\newblock


\bibitem[Lentz et~al\mbox{.}(2007)]%
        {dynamic1}
\bibfield{author}{\bibinfo{person}{Tobias Lentz}, \bibinfo{person}{Dirk
  Schr\"{o}der}, \bibinfo{person}{Michael Vorl\"{a}nder}, {and}
  \bibinfo{person}{Ingo Assenmacher}.} \bibinfo{year}{2007}\natexlab{}.
\newblock \showarticletitle{Virtual Reality System with Integrated Sound Field
  Simulation and Reproduction}.
\newblock \bibinfo{journal}{\emph{EURASIP J. Adv. Signal Process}}
  \bibinfo{volume}{2007}, \bibinfo{number}{1} (\bibinfo{date}{jan}
  \bibinfo{year}{2007}), \bibinfo{pages}{187}.
\newblock
\showISSN{1110-8657}
\urldef\tempurl%
\url{https://doi.org/10.1155/2007/70540}
\showDOI{\tempurl}


\bibitem[Liu and Manocha(2020)]%
        {SURVEY_SOUND}
\bibfield{author}{\bibinfo{person}{Shiguang Liu} {and} \bibinfo{person}{Dinesh
  Manocha}.} \bibinfo{year}{2020}\natexlab{}.
\newblock \showarticletitle{Sound Synthesis, Propagation, and Rendering: {A}
  Survey}.
\newblock \bibinfo{journal}{\emph{CoRR}}  \bibinfo{volume}{abs/2011.05538}
  (\bibinfo{year}{2020}).
\newblock


\bibitem[Luo et~al\mbox{.}(2022)]%
        {neural_acoustic_field}
\bibfield{author}{\bibinfo{person}{Andrew Luo}, \bibinfo{person}{Yilun Du},
  \bibinfo{person}{Michael~J. Tarr}, \bibinfo{person}{Joshua~B. Tenenbaum},
  \bibinfo{person}{Antonio Torralba}, {and} \bibinfo{person}{Chuang Gan}.}
  \bibinfo{year}{2022}\natexlab{}.
\newblock \showarticletitle{Learning Neural Acoustic Fields}.
\newblock \bibinfo{journal}{\emph{CoRR}}  \bibinfo{volume}{abs/2204.00628}
  (\bibinfo{year}{2022}).
\newblock


\bibitem[Luo et~al\mbox{.}(2020)]%
        {luo2020dual}
\bibfield{author}{\bibinfo{person}{Yi Luo}, \bibinfo{person}{Zhuo Chen}, {and}
  \bibinfo{person}{Takuya Yoshioka}.} \bibinfo{year}{2020}\natexlab{}.
\newblock \showarticletitle{Dual-path rnn: efficient long sequence modeling for
  time-domain single-channel speech separation}. In
  \bibinfo{booktitle}{\emph{ICASSP 2020-2020 IEEE International Conference on
  Acoustics, Speech and Signal Processing (ICASSP)}}. IEEE,
  \bibinfo{pages}{46--50}.
\newblock


\bibitem[Malik et~al\mbox{.}(2021)]%
        {speech_recognition1}
\bibfield{author}{\bibinfo{person}{Mishaim Malik},
  \bibinfo{person}{Muhammad~Kamran Malik}, \bibinfo{person}{Khawar Mehmood},
  {and} \bibinfo{person}{Imran Makhdoom}.} \bibinfo{year}{2021}\natexlab{}.
\newblock \showarticletitle{Automatic speech recognition: a survey}.
\newblock \bibinfo{journal}{\emph{Multim. Tools Appl.}} \bibinfo{volume}{80},
  \bibinfo{number}{6} (\bibinfo{year}{2021}), \bibinfo{pages}{9411--9457}.
\newblock


\bibitem[Mehra et~al\mbox{.}(2013)]%
        {precompute3}
\bibfield{author}{\bibinfo{person}{Ravish Mehra}, \bibinfo{person}{Nikunj
  Raghuvanshi}, \bibinfo{person}{Lakulish Antani}, \bibinfo{person}{Anish
  Chandak}, \bibinfo{person}{Sean Curtis}, {and} \bibinfo{person}{Dinesh
  Manocha}.} \bibinfo{year}{2013}\natexlab{}.
\newblock \showarticletitle{Wave-Based Sound Propagation in Large Open Scenes
  Using an Equivalent Source Formulation}.
\newblock \bibinfo{journal}{\emph{ACM Trans. Graph.}} \bibinfo{volume}{32},
  \bibinfo{number}{2}, Article \bibinfo{articleno}{19} (\bibinfo{date}{apr}
  \bibinfo{year}{2013}), \bibinfo{numpages}{13}~pages.
\newblock
\showISSN{0730-0301}
\urldef\tempurl%
\url{https://doi.org/10.1145/2451236.2451245}
\showDOI{\tempurl}


\bibitem[Meng et~al\mbox{.}(2021)]%
        {finite2}
\bibfield{author}{\bibinfo{person}{Hsien{-}Yu Meng}, \bibinfo{person}{Zhenyu
  Tang}, {and} \bibinfo{person}{Dinesh Manocha}.}
  \bibinfo{year}{2021}\natexlab{}.
\newblock \showarticletitle{Point-based Acoustic Scattering for Interactive
  Sound Propagation via Surface Encoding}. In
  \bibinfo{booktitle}{\emph{{IJCAI}}}. \bibinfo{publisher}{ijcai.org},
  \bibinfo{pages}{909--915}.
\newblock


\bibitem[Mirza and Osindero(2014)]%
        {CGAN1}
\bibfield{author}{\bibinfo{person}{Mehdi Mirza} {and} \bibinfo{person}{Simon
  Osindero}.} \bibinfo{year}{2014}\natexlab{}.
\newblock \showarticletitle{Conditional Generative Adversarial Nets}.
\newblock \bibinfo{journal}{\emph{arXiv preprint arXiv:1411.1784}}
  (\bibinfo{year}{2014}).
\newblock


\bibitem[Muntoni and Cignoni(2021)]%
        {pymeshlab}
\bibfield{author}{\bibinfo{person}{Alessandro Muntoni} {and}
  \bibinfo{person}{Paolo Cignoni}.} \bibinfo{year}{2021}\natexlab{}.
\newblock \bibinfo{booktitle}{\emph{{PyMeshLab}}}.
\newblock
\urldef\tempurl%
\url{https://doi.org/10.5281/zenodo.4438750}
\showDOI{\tempurl}


\bibitem[Nakamura et~al\mbox{.}(1999)]%
        {nakamura1999sound}
\bibfield{author}{\bibinfo{person}{Satoshi Nakamura}, \bibinfo{person}{Kazuo
  Hiyane}, \bibinfo{person}{Futoshi Asano}, {and} \bibinfo{person}{Takashi
  Endo}.} \bibinfo{year}{1999}\natexlab{}.
\newblock \showarticletitle{Sound scene data collection in real acoustical
  environments}.
\newblock \bibinfo{journal}{\emph{Journal of the Acoustical Society of Japan
  (E)}} \bibinfo{volume}{20}, \bibinfo{number}{3} (\bibinfo{year}{1999}),
  \bibinfo{pages}{225--231}.
\newblock


\bibitem[Naylor and Gaubitch(2010)]%
        {drr_book}
\bibfield{author}{\bibinfo{person}{P~A Naylor} {and} \bibinfo{person}{N~D
  Gaubitch}.} \bibinfo{year}{2010}\natexlab{}.
\newblock \bibinfo{booktitle}{\emph{Speech Dereverberation}
  (\bibinfo{edition}{1st} ed.)}.
\newblock \bibinfo{publisher}{Springer Publishing Company, Incorporated}.
\newblock
\showISBNx{1849960550}


\bibitem[Neo et~al\mbox{.}(2020)]%
        {speech_enhancement}
\bibfield{author}{\bibinfo{person}{Vincent~W. Neo}, \bibinfo{person}{Christine
  Evers}, {and} \bibinfo{person}{Patrick~A. Naylor}.}
  \bibinfo{year}{2020}\natexlab{}.
\newblock \showarticletitle{PEVD-Based Speech Enhancement in Reverberant
  Environments}. In \bibinfo{booktitle}{\emph{{ICASSP}}}.
  \bibinfo{publisher}{{IEEE}}, \bibinfo{pages}{186--190}.
\newblock


\bibitem[Owens et~al\mbox{.}(2016)]%
        {video2audio}
\bibfield{author}{\bibinfo{person}{Andrew Owens}, \bibinfo{person}{Phillip
  Isola}, \bibinfo{person}{Josh~H. McDermott}, \bibinfo{person}{Antonio
  Torralba}, \bibinfo{person}{Edward~H. Adelson}, {and}
  \bibinfo{person}{William~T. Freeman}.} \bibinfo{year}{2016}\natexlab{}.
\newblock \showarticletitle{Visually Indicated Sounds}. In
  \bibinfo{booktitle}{\emph{{CVPR}}}. \bibinfo{publisher}{{IEEE} Computer
  Society}, \bibinfo{pages}{2405--2413}.
\newblock


\bibitem[Pariente et~al\mbox{.}(2020)]%
        {pariente2020asteroid}
\bibfield{author}{\bibinfo{person}{Manuel Pariente}, \bibinfo{person}{Samuele
  Cornell}, \bibinfo{person}{Joris Cosentino}, \bibinfo{person}{Sunit
  Sivasankaran}, \bibinfo{person}{Efthymios Tzinis}, \bibinfo{person}{Jens
  Heitkaemper}, \bibinfo{person}{Michel Olvera}, \bibinfo{person}{Fabian-Robert
  St{\"o}ter}, \bibinfo{person}{Mathieu Hu}, \bibinfo{person}{Juan~M
  Mart{\'\i}n-Do{\~n}as}, {et~al\mbox{.}}} \bibinfo{year}{2020}\natexlab{}.
\newblock \showarticletitle{Asteroid: the PyTorch-based audio source separation
  toolkit for researchers}.
\newblock \bibinfo{journal}{\emph{arXiv preprint arXiv:2005.04132}}
  (\bibinfo{year}{2020}).
\newblock


\bibitem[Pulkki and Svensson(2019)]%
        {finite1}
\bibfield{author}{\bibinfo{person}{Ville Pulkki} {and}
  \bibinfo{person}{U.~Peter Svensson}.} \bibinfo{year}{2019}\natexlab{}.
\newblock \showarticletitle{Machine-learning-based estimation and rendering of
  scattering in virtual reality}.
\newblock \bibinfo{journal}{\emph{The Journal of the Acoustical Society of
  America}} \bibinfo{volume}{145}, \bibinfo{number}{4} (\bibinfo{year}{2019}),
  \bibinfo{pages}{2664--2676}.
\newblock
\urldef\tempurl%
\url{https://doi.org/10.1121/1.5095875}
\showDOI{\tempurl}
\showeprint{https://doi.org/10.1121/1.5095875}


\bibitem[Raghuvanshi et~al\mbox{.}(2010)]%
        {precompute2}
\bibfield{author}{\bibinfo{person}{Nikunj Raghuvanshi}, \bibinfo{person}{John
  Snyder}, \bibinfo{person}{Ravish Mehra}, \bibinfo{person}{Ming Lin}, {and}
  \bibinfo{person}{Naga Govindaraju}.} \bibinfo{year}{2010}\natexlab{}.
\newblock \showarticletitle{Precomputed Wave Simulation for Real-Time Sound
  Propagation of Dynamic Sources in Complex Scenes}.
\newblock \bibinfo{journal}{\emph{ACM Trans. Graph.}} \bibinfo{volume}{29},
  \bibinfo{number}{4}, Article \bibinfo{articleno}{68} (\bibinfo{date}{jul}
  \bibinfo{year}{2010}), \bibinfo{numpages}{11}~pages.
\newblock
\showISSN{0730-0301}
\urldef\tempurl%
\url{https://doi.org/10.1145/1778765.1778805}
\showDOI{\tempurl}


\bibitem[Ratnarajah et~al\mbox{.}(2021a)]%
        {ir-gan}
\bibfield{author}{\bibinfo{person}{Anton Ratnarajah}, \bibinfo{person}{Zhenyu
  Tang}, {and} \bibinfo{person}{Dinesh Manocha}.}
  \bibinfo{year}{2021}\natexlab{a}.
\newblock \showarticletitle{{IR-GAN: Room Impulse Response Generator for
  Far-Field Speech Recognition}}. In \bibinfo{booktitle}{\emph{Proc.
  Interspeech 2021}}. \bibinfo{pages}{286--290}.
\newblock
\urldef\tempurl%
\url{https://doi.org/10.21437/Interspeech.2021-230}
\showDOI{\tempurl}


\bibitem[Ratnarajah et~al\mbox{.}(2021b)]%
        {ts-rir}
\bibfield{author}{\bibinfo{person}{Anton Ratnarajah}, \bibinfo{person}{Zhenyu
  Tang}, {and} \bibinfo{person}{Dinesh Manocha}.}
  \bibinfo{year}{2021}\natexlab{b}.
\newblock \showarticletitle{{TS-RIR:} Translated Synthetic Room Impulse
  Responses for Speech Augmentation}. In \bibinfo{booktitle}{\emph{{ASRU}}}.
  \bibinfo{publisher}{{IEEE}}, \bibinfo{pages}{259--266}.
\newblock


\bibitem[Ratnarajah et~al\mbox{.}(2021c)]%
        {fast-rir}
\bibfield{author}{\bibinfo{person}{Anton Ratnarajah},
  \bibinfo{person}{Shi{-}Xiong Zhang}, \bibinfo{person}{Meng Yu},
  \bibinfo{person}{Zhenyu Tang}, \bibinfo{person}{Dinesh Manocha}, {and}
  \bibinfo{person}{Dong Yu}.} \bibinfo{year}{2021}\natexlab{c}.
\newblock \showarticletitle{{FAST-RIR:} Fast neural diffuse room impulse
  response generator}.
\newblock \bibinfo{journal}{\emph{CoRR}}  \bibinfo{volume}{abs/2110.04057}
  (\bibinfo{year}{2021}).
\newblock


\bibitem[Richard et~al\mbox{.}(2022)]%
        {deepIR}
\bibfield{author}{\bibinfo{person}{Alexander Richard}, \bibinfo{person}{Peter
  Dodds}, {and} \bibinfo{person}{Vamsi~Krishna Ithapu}.}
  \bibinfo{year}{2022}\natexlab{}.
\newblock \showarticletitle{Deep Impulse Responses: Estimating and
  Parameterizing Filters with Deep Networks}.
\newblock \bibinfo{journal}{\emph{CoRR}}  \bibinfo{volume}{abs/2202.03416}
  (\bibinfo{year}{2022}).
\newblock


\bibitem[Richey et~al\mbox{.}(2018)]%
        {richey2018voices}
\bibfield{author}{\bibinfo{person}{Colleen Richey}, \bibinfo{person}{Maria~A
  Barrios}, \bibinfo{person}{Zeb Armstrong}, \bibinfo{person}{Chris Bartels},
  \bibinfo{person}{Horacio Franco}, \bibinfo{person}{Martin Graciarena},
  \bibinfo{person}{Aaron Lawson}, \bibinfo{person}{Mahesh~Kumar Nandwana},
  \bibinfo{person}{Allen Stauffer}, \bibinfo{person}{Julien van Hout},
  {et~al\mbox{.}}} \bibinfo{year}{2018}\natexlab{}.
\newblock \showarticletitle{Voices obscured in complex environmental settings
  (voices) corpus}.
\newblock \bibinfo{journal}{\emph{arXiv preprint arXiv:1804.05053}}
  (\bibinfo{year}{2018}).
\newblock


\bibitem[Schissler and Manocha(2011)]%
        {gsound}
\bibfield{author}{\bibinfo{person}{Carl Schissler} {and}
  \bibinfo{person}{Dinesh Manocha}.} \bibinfo{year}{2011}\natexlab{}.
\newblock \showarticletitle{{GSound: Interactive Sound Propagation for Games}}.
\newblock \bibinfo{journal}{\emph{Journal of the audio engineering society}}
  (\bibinfo{date}{February} \bibinfo{year}{2011}).
\newblock


\bibitem[Schissler and Manocha(2016)]%
        {dynamic3}
\bibfield{author}{\bibinfo{person}{Carl Schissler} {and}
  \bibinfo{person}{Dinesh Manocha}.} \bibinfo{year}{2016}\natexlab{}.
\newblock \showarticletitle{Interactive Sound Propagation and Rendering for
  Large Multi-Source Scenes}.
\newblock \bibinfo{journal}{\emph{ACM Trans. Graph.}} \bibinfo{volume}{36},
  \bibinfo{number}{4}, Article \bibinfo{articleno}{114c} (\bibinfo{date}{sep}
  \bibinfo{year}{2016}), \bibinfo{numpages}{12}~pages.
\newblock
\showISSN{0730-0301}
\urldef\tempurl%
\url{https://doi.org/10.1145/3072959.2943779}
\showDOI{\tempurl}


\bibitem[Schissler et~al\mbox{.}(2014)]%
        {source1}
\bibfield{author}{\bibinfo{person}{Carl Schissler}, \bibinfo{person}{Ravish
  Mehra}, {and} \bibinfo{person}{Dinesh Manocha}.}
  \bibinfo{year}{2014}\natexlab{}.
\newblock \showarticletitle{High-Order Diffraction and Diffuse Reflections for
  Interactive Sound Propagation in Large Environments}.
\newblock \bibinfo{journal}{\emph{ACM Trans. Graph.}} \bibinfo{volume}{33},
  \bibinfo{number}{4}, Article \bibinfo{articleno}{39} (\bibinfo{date}{jul}
  \bibinfo{year}{2014}), \bibinfo{numpages}{12}~pages.
\newblock
\showISSN{0730-0301}
\urldef\tempurl%
\url{https://doi.org/10.1145/2601097.2601216}
\showDOI{\tempurl}


\bibitem[Schroeder(1965)]%
        {EDC}
\bibfield{author}{\bibinfo{person}{M.~R. Schroeder}.}
  \bibinfo{year}{1965}\natexlab{}.
\newblock \showarticletitle{New Method of Measuring Reverberation Time}.
\newblock \bibinfo{journal}{\emph{The Journal of the Acoustical Society of
  America}} \bibinfo{volume}{37}, \bibinfo{number}{6} (\bibinfo{year}{1965}),
  \bibinfo{pages}{1187--1188}.
\newblock
\urldef\tempurl%
\url{https://doi.org/10.1121/1.1939454}
\showDOI{\tempurl}
\showeprint{https://doi.org/10.1121/1.1939454}


\bibitem[Shao et~al\mbox{.}(2021)]%
        {speech_recognition2}
\bibfield{author}{\bibinfo{person}{Yiwen Shao}, \bibinfo{person}{Shi{-}Xiong
  Zhang}, {and} \bibinfo{person}{Dong Yu}.} \bibinfo{year}{2021}\natexlab{}.
\newblock \showarticletitle{Multi-Channel Multi-Speaker {ASR} Using 3D Spatial
  Feature}.
\newblock \bibinfo{journal}{\emph{CoRR}}  \bibinfo{volume}{abs/2111.11023}
  (\bibinfo{year}{2021}).
\newblock


\bibitem[Singh et~al\mbox{.}(2021)]%
        {image2reverb}
\bibfield{author}{\bibinfo{person}{Nikhil Singh}, \bibinfo{person}{Jeff
  Mentch}, \bibinfo{person}{Jerry Ng}, \bibinfo{person}{Matthew Beveridge},
  {and} \bibinfo{person}{Iddo Drori}.} \bibinfo{year}{2021}\natexlab{}.
\newblock \showarticletitle{Image2Reverb: Cross-Modal Reverb Impulse Response
  Synthesis}. In \bibinfo{booktitle}{\emph{{ICCV}}}.
  \bibinfo{publisher}{{IEEE}}, \bibinfo{pages}{286--295}.
\newblock


\bibitem[Steinmetz et~al\mbox{.}(2021)]%
        {ir_estimate}
\bibfield{author}{\bibinfo{person}{Christian~J. Steinmetz},
  \bibinfo{person}{Vamsi~Krishna Ithapu}, {and} \bibinfo{person}{Paul
  Calamia}.} \bibinfo{year}{2021}\natexlab{}.
\newblock \showarticletitle{Filtered Noise Shaping for Time Domain Room Impulse
  Response Estimation from Reverberant Speech}. In
  \bibinfo{booktitle}{\emph{{WASPAA}}}. \bibinfo{publisher}{{IEEE}},
  \bibinfo{pages}{221--225}.
\newblock


\bibitem[Sz{\"{o}}ke et~al\mbox{.}(2019)]%
        {realIR2}
\bibfield{author}{\bibinfo{person}{Igor Sz{\"{o}}ke}, \bibinfo{person}{Miroslav
  Sk{\'{a}}cel}, \bibinfo{person}{Ladislav Mosner}, \bibinfo{person}{Jakub
  Paliesek}, {and} \bibinfo{person}{Jan~Honza Cernock{\'{y}}}.}
  \bibinfo{year}{2019}\natexlab{}.
\newblock \showarticletitle{Building and Evaluation of a Real Room Impulse
  Response Dataset}.
\newblock \bibinfo{journal}{\emph{{IEEE} J. Sel. Top. Signal Process.}}
  \bibinfo{volume}{13}, \bibinfo{number}{4} (\bibinfo{year}{2019}),
  \bibinfo{pages}{863--876}.
\newblock


\bibitem[Szöke et~al\mbox{.}(2019)]%
        {8717722}
\bibfield{author}{\bibinfo{person}{Igor Szöke}, \bibinfo{person}{Miroslav
  Skácel}, \bibinfo{person}{Ladislav Mošner}, \bibinfo{person}{Jakub
  Paliesek}, {and} \bibinfo{person}{Jan Černocký}.}
  \bibinfo{year}{2019}\natexlab{}.
\newblock \showarticletitle{Building and evaluation of a real room impulse
  response dataset}.
\newblock \bibinfo{journal}{\emph{IEEE Journal of Selected Topics in Signal
  Processing}} \bibinfo{volume}{13}, \bibinfo{number}{4}
  (\bibinfo{year}{2019}), \bibinfo{pages}{863--876}.
\newblock
\urldef\tempurl%
\url{https://doi.org/10.1109/JSTSP.2019.2917582}
\showDOI{\tempurl}


\bibitem[Tang et~al\mbox{.}(2022)]%
        {GWA}
\bibfield{author}{\bibinfo{person}{Zhenyu Tang}, \bibinfo{person}{Rohith
  Aralikatti}, \bibinfo{person}{Anton Ratnarajah}, {and}
  \bibinfo{person}{Dinesh Manocha}.} \bibinfo{year}{2022}\natexlab{}.
\newblock \showarticletitle{GWA: A Large High-Quality Acoustic Dataset for
  Audio Processing}.
\newblock \bibinfo{journal}{\emph{ArXiv}}  \bibinfo{volume}{abs/2204.01787}
  (\bibinfo{year}{2022}).
\newblock


\bibitem[Tang et~al\mbox{.}(2020)]%
        {pygsound}
\bibfield{author}{\bibinfo{person}{Zhenyu Tang}, \bibinfo{person}{Lianwu Chen},
  \bibinfo{person}{Bo Wu}, \bibinfo{person}{Dong Yu}, {and}
  \bibinfo{person}{Dinesh Manocha}.} \bibinfo{year}{2020}\natexlab{}.
\newblock \showarticletitle{Improving Reverberant Speech Training Using Diffuse
  Acoustic Simulation}. In \bibinfo{booktitle}{\emph{{ICASSP}}}.
  \bibinfo{publisher}{{IEEE}}, \bibinfo{pages}{6969--6973}.
\newblock


\bibitem[Taylor et~al\mbox{.}(2012)]%
        {dynamic2}
\bibfield{author}{\bibinfo{person}{Micah Taylor}, \bibinfo{person}{Anish
  Chandak}, \bibinfo{person}{Qi Mo}, \bibinfo{person}{Christian Lauterbach},
  \bibinfo{person}{Carl Schissler}, {and} \bibinfo{person}{Dinesh Manocha}.}
  \bibinfo{year}{2012}\natexlab{}.
\newblock \showarticletitle{Guided Multiview Ray Tracing for Fast
  Auralization}.
\newblock \bibinfo{journal}{\emph{IEEE Transactions on Visualization and
  Computer Graphics}} \bibinfo{volume}{18}, \bibinfo{number}{11}
  (\bibinfo{year}{2012}), \bibinfo{pages}{1797--1810}.
\newblock
\urldef\tempurl%
\url{https://doi.org/10.1109/TVCG.2012.27}
\showDOI{\tempurl}


\bibitem[Traer and McDermott(2016)]%
        {traer2016statistics}
\bibfield{author}{\bibinfo{person}{James Traer} {and} \bibinfo{person}{Josh~H
  McDermott}.} \bibinfo{year}{2016}\natexlab{}.
\newblock \showarticletitle{Statistics of natural reverberation enable
  perceptual separation of sound and space}.
\newblock \bibinfo{journal}{\emph{Proceedings of the National Academy of
  Sciences}} \bibinfo{volume}{113}, \bibinfo{number}{48}
  (\bibinfo{year}{2016}), \bibinfo{pages}{E7856--E7865}.
\newblock


\bibitem[Wang et~al\mbox{.}(2018)]%
        {image2mesh}
\bibfield{author}{\bibinfo{person}{Nanyang Wang}, \bibinfo{person}{Yinda
  Zhang}, \bibinfo{person}{Zhuwen Li}, \bibinfo{person}{Yanwei Fu},
  \bibinfo{person}{Wei Liu}, {and} \bibinfo{person}{Yu{-}Gang Jiang}.}
  \bibinfo{year}{2018}\natexlab{}.
\newblock \showarticletitle{Pixel2Mesh: Generating 3D Mesh Models from Single
  {RGB} Images}. In \bibinfo{booktitle}{\emph{{ECCV} {(11)}}}
  \emph{(\bibinfo{series}{Lecture Notes in Computer Science},
  Vol.~\bibinfo{volume}{11215})}. \bibinfo{publisher}{Springer},
  \bibinfo{pages}{55--71}.
\newblock


\bibitem[Werner and Liebetrau(2014)]%
        {werner2014adjustment}
\bibfield{author}{\bibinfo{person}{Stephan Werner} {and}
  \bibinfo{person}{Judith Liebetrau}.} \bibinfo{year}{2014}\natexlab{}.
\newblock \showarticletitle{Adjustment of direct-to-reverberant-energy-ratio
  and the just-noticable-difference}. In \bibinfo{booktitle}{\emph{2014 Sixth
  International Workshop on Quality of Multimedia Experience (QoMEX)}}. IEEE,
  \bibinfo{pages}{1--3}.
\newblock


\bibitem[Wrobel(2003)]%
        {boundary-element}
\bibfield{author}{\bibinfo{person}{LC Wrobel}.}
  \bibinfo{year}{2003}\natexlab{}.
\newblock \showarticletitle{{Boundary Element Method, Volume 1: Applications in
  Thermo-Fluids and Acoustics}}.
\newblock \bibinfo{journal}{\emph{Applied Mechanics Reviews}}
  \bibinfo{volume}{56}, \bibinfo{number}{2} (\bibinfo{date}{03}
  \bibinfo{year}{2003}), \bibinfo{pages}{B17--B17}.
\newblock
\showISSN{0003-6900}
\urldef\tempurl%
\url{https://doi.org/10.1115/1.1553431}
\showDOI{\tempurl}


\bibitem[Yeh et~al\mbox{.}(2013)]%
        {precompute1}
\bibfield{author}{\bibinfo{person}{Hengchin Yeh}, \bibinfo{person}{Ravish
  Mehra}, \bibinfo{person}{Zhimin Ren}, \bibinfo{person}{Lakulish Antani},
  \bibinfo{person}{Dinesh Manocha}, {and} \bibinfo{person}{Ming Lin}.}
  \bibinfo{year}{2013}\natexlab{}.
\newblock \showarticletitle{Wave-Ray Coupling for Interactive Sound Propagation
  in Large Complex Scenes}.
\newblock \bibinfo{journal}{\emph{ACM Trans. Graph.}} \bibinfo{volume}{32},
  \bibinfo{number}{6}, Article \bibinfo{articleno}{165} (\bibinfo{date}{nov}
  \bibinfo{year}{2013}), \bibinfo{numpages}{11}~pages.
\newblock
\showISSN{0730-0301}
\urldef\tempurl%
\url{https://doi.org/10.1145/2508363.2508420}
\showDOI{\tempurl}


\bibitem[Zhang et~al\mbox{.}(2017)]%
        {stackgan}
\bibfield{author}{\bibinfo{person}{Han Zhang}, \bibinfo{person}{Tao Xu},
  \bibinfo{person}{Hongsheng Li}, \bibinfo{person}{Shaoting Zhang},
  \bibinfo{person}{Xiaogang Wang}, \bibinfo{person}{Xiaolei Huang}, {and}
  \bibinfo{person}{Dimitris Metaxas}.} \bibinfo{year}{2017}\natexlab{}.
\newblock \showarticletitle{StackGAN: Text to Photo-realistic Image Synthesis
  with Stacked Generative Adversarial Networks}. In
  \bibinfo{booktitle}{\emph{{ICCV}}}.
\newblock


\bibitem[Zhang et~al\mbox{.}(2019)]%
        {stackgan++}
\bibfield{author}{\bibinfo{person}{Han Zhang}, \bibinfo{person}{Tao Xu},
  \bibinfo{person}{Hongsheng Li}, \bibinfo{person}{Shaoting Zhang},
  \bibinfo{person}{Xiaogang Wang}, \bibinfo{person}{Xiaolei Huang}, {and}
  \bibinfo{person}{Dimitris~N. Metaxas}.} \bibinfo{year}{2019}\natexlab{}.
\newblock \showarticletitle{StackGAN++: Realistic Image Synthesis with Stacked
  Generative Adversarial Networks}.
\newblock \bibinfo{journal}{\emph{{IEEE} Trans. Pattern Anal. Mach. Intell.}}
  \bibinfo{volume}{41}, \bibinfo{number}{8} (\bibinfo{year}{2019}),
  \bibinfo{pages}{1947--1962}.
\newblock


\end{thebibliography}

\end{document}